\documentclass[a4paper,11pt]{article}

\usepackage{amsmath}
\usepackage{amsfonts}
\usepackage{amssymb}
\usepackage{textcomp}
\usepackage{graphicx}
\usepackage{bm}
\usepackage{color}
\usepackage{verbatim}
\usepackage{subfig}
\usepackage{hyperref}

\usepackage{amsfonts,amsmath,amssymb}

\newcommand{\be}{\begin{equation}}
\newcommand{\ee}{\end{equation}}
\newcommand{\ba}{\begin{eqnarray}}
\newcommand{\ea}{\end{eqnarray}}
\newcommand{\ben}{\begin{enumerate}}
\newcommand{\een}{\end{enumerate}}

\newcommand{\de}{\delta}
\newcommand{\ep}{\epsilon}

\newcommand{\p}{\partial}

\newcommand{\la}{\langle}
\newcommand{\ra}{\rangle}

\newcommand{\rar}{\rightarrow}
\newcommand{\eps}{\epsilon}
\newcommand{\bom}{\bar{\omega}}

\newcommand{\beq}{\begin{eqnarray}}
\newcommand{\eeq}{\end{eqnarray}}
\newcommand{\non}{\nonumber\\}

\renewcommand{\Im}{\mathop{\rm Im}}

\begin{document}

\begin{titlepage}
\def\thefootnote{\fnsymbol{footnote}}
\phantom{.}\vspace{-2cm}
\begin{flushright}\footnotesize
\texttt{NORDITA-2014-61} \\
\texttt{UUITP-04/14}
\vspace{0.6cm}
\end{flushright}

\bigskip

\begin{center}
{\Large {\bf Classical and quantum temperature fluctuations via holography}}

\bigskip
{\large 
Alexander Balatsky${}^{1,2}$,
Sven Bjarke Gudnason${}^1$,
Yaron Kedem${}^1$,
Alexander Krikun${}^{1,3}$,\\
L\'arus Thorlacius${}^{1,4,5}$
and Konstantin Zarembo${}^{1,3,6}$
}
\end{center}

\renewcommand{\thefootnote}{\arabic{footnote}}

\begin{center}
\vspace{0em}
{\em ${}^1$Nordita, KTH Royal Institute of Technology and Stockholm University,
Roslagstullsbacken 23, SE-106 91 Stockholm, Sweden\\
${}^2$Institute for Materials Science, Los Alamos National Laboratory,
Los Alamos, NM 87545, USA\\
${}^3$Institute of Theoretical and Experimental Physics,
B.~Cheremushkinskaya 25, 117218 Moscow, Russia\\
${}^4$University of Iceland, Science Institute, Dunhaga 3, IS-107
Reykjavik, Iceland\\
${}^5$ The Oskar Klein Centre for Cosmoparticle Physics, Department of Physics, Stockholm University,
AlbaNova University Centre, 10691 Stockholm, Sweden\\
${}^6$Department of Physics and Astronomy, Uppsala University
SE-751 08 Uppsala, Sweden\\

\vskip .4cm}

\end{center}

\vspace{1.1cm}

\noindent
\begin{center} {\bf Abstract} \end{center}
We study local temperature fluctuations in a 2+1 dimensional CFT
on the sphere, dual to a black hole in asymptotically AdS
spacetime. 
The fluctuation spectrum is governed by the lowest-lying hydrodynamic
modes of the system whose frequency and damping rate determine
whether temperature fluctuations are thermal or quantum. 
We calculate numerically the corresponding quasinormal frequencies and
match the result with the hydrodynamics of the dual CFT at large 
temperature.  
As a by-product of our analysis we determine the appropriate boundary
conditions for calculating low-lying quasinormal modes for a
four-dimensional Reissner-Nordstr\"om  black hole in global AdS. 
\vfill

\begin{flushleft}
{\today}
\end{flushleft}
\end{titlepage}

\hfill{}
\setcounter{footnote}{0}

\section{Introduction}

The field of black hole thermodynamics was cast into life by the
definition of black hole entropy and temperature in the seminal papers
by Bekenstein \cite{Bekenstein:1973ur} and Hawking
\cite{Hawking:1974sw}. Early on it was viewed as an interesting analogy
between dynamical equations of two rather different fields of physics,
general relativity and statistical mechanics, but with the advent of 
anti de Sitter - conformal field theory (AdS/CFT) duality
\cite{Maldacena:1997re} the analogy has been promoted to a precise
correspondence for a class of black holes. 

A Schwarzschild black hole in asymptotically flat spacetime is unstable due to 
the Hawking effect. It evaporates if it is surrounded by empty spacetime and 
as it has negative heat capacity it cannot be in stable equilibrium with a
thermal gas either. In fact, due to the Jeans instability, the thermodynamic
limit is not well defined in Einstein gravity in asymptotically flat spacetime
and there exist no equilibrium configurations at finite temperature and 
density. 

The situation is different in asymptotically AdS spacetime, where a large black 
hole above the Hawking-Page phase transition \cite{Hawking:1982dh} is a 
stable configuration. Under AdS/CFT duality, such a black hole corresponds 
to a thermal state in the conformal field theory at the boundary of spacetime,
with the CFT temperature equal to the Hawking temperature of the black 
hole \cite{Witten:1998zw}. The appropriate interpretation of black hole
thermodynamics in asymptotically AdS spacetime is in terms of the dual 
field theory rather than spacetime physics in the bulk. The AdS black hole 
geometry itself is a solution of the classical field equations of
general relativity without matter and observers in free fall outside a
large AdS black hole do not detect any propagating Hawking radiation
\cite{Brynjolfsson:2008uc}. Nonetheless the Hawking temperature,
interpreted as the temperature of the dual CFT, is a well-defined observable.

Any macroscopic physical observable is subject to statistical
fluctuations. When one measures a particular quantity one gets results
which are distributed around the mean value with some finite standard
deviation. For macroscopic objects the fluctuations are usually
governed by random thermal noise, but when the system is small enough,
the quantum uncertainty principle starts playing a significant role
and changes the overall behavior of the system
\cite{balatsky2003quantum}. Thus one distinguishes between the regimes
of thermal and quantum fluctuations of the observable. This treatment
can readily be applied to fluctuations in the temperature of the
system. Due to energy conservation, the mean value of the temperature
in the dual field theory does not fluctuate with time. However, local
temperature fluctuations do exist and they will exhibit thermal or
quantum behavior depending on system parameters. 

In this paper we use the AdS/CFT duality to investigate local
temperature fluctuations in a field-theoretic system at strong
coupling via the dynamics of asymptotically AdS black holes. We focus
on the AdS-Reissner-Nordstr\"om solution dual to a CFT at finite
temperature and chemical potential. The rationale for switching on a
chemical potential is that it allows us to consider low temperature
without compromising thermodynamic stability, which as we shall see
leads to a crossover from a thermal to a quantum regime in temperature 
fluctuations. 

The paper is organized as follows. In Section~\ref{Sec:1} we recall 
the notion of temperature fluctuations and analyze them in the case of
overdamped and underdamped modes.
In Section~\ref{Sec:2} we set up the problem of calculating local
temperature fluctuations of the CFT in compact space. Sections~\ref{Sec:Hydro} 
and~\ref{Sec:charge} consider  
the hydrodynamic approximation to the sound and diffusive modes of the
CFT on a sphere, dual to the black hole under consideration, in case
of zero and finite chemical potential, respectively. In Section~\ref{Sec:3} 
we carry out a direct gravitational analysis of the system
and compute numerically the lowest-lying quasinormal modes of the
black hole. We conclude in Section~\ref{Conclusion}. 
Appendix~\ref{app:A} is devoted to a detailed analysis of the
transition between classical and quantum regimes of the fluctuations. 
Appendix~\ref{Thermo} collects various thermodynamic relations, that we
use throughout the paper. 
In Appendix~\ref{app:B} we consider details of the calculation of QNMs 
of the Reissner-Nordstr\"om black hole in global AdS${}_4$.

\section{\label{Sec:1} Temperature fluctuations}

Temperature fluctuations in classical systems have been widely
studied. In classical statistical mechanics the variance of a
statistical variable is given by the width of its probability
distribution, which for temperature gives
\cite{lifshitz1984statistical}\footnote{We set $k_B=1$ but will keep 
the explicit dependence on $\hbar$ in this section and in
App.~\ref{app:A} for distinguishing classical and quantum
contributions. } 
\beq
\langle\delta T^2\rangle = \frac{T^2}{C_v},
\label{eq:classical_fluctuations}
\eeq
where $T$ is the temperature and $C_v$ is the heat capacity. From this 
expression, it is clear that in order for the temperature to be a
well-defined quantity, the heat capacity must be large. 

Temperature fluctuations close to equilibrium can be described by
linear response theory \cite{Nyquist:1928zz,balatsky2003quantum}. To
this end, we consider a system in which the temperature of a black
body is in equilibrium with the surrounding radiation. In response to
an external perturbation, the system will relax to equilibrium on a
characteristic time-scale $\tau$, with its temperature changing with
time according to 
\beq
\frac{dT}{dt} = -\frac{T-T_e}{\tau},
\label{eq:dTdt}
\eeq
with $T_e$ being the equilibrium temperature. The change in
entropy $\Delta S$ caused by a perturbation is related to the change
in temperature as 
\beq
\Delta T_e = \frac{\p T}{\p S}\Delta S = \frac{T}{C_v} \Delta S.
\label{eq:deltaT_e} 
\eeq
The Fourier spectrum of temperature fluctuations is thus related to
that of the entropy by a response function (generalized
susceptibility): 
\beq
\delta T(\omega) = \alpha(\omega) \Delta S(\omega),
\eeq
given by \eqref{eq:deltaT_e}
and \eqref{eq:dTdt}:
\beq
\alpha(\omega) = \frac{T}{C_v}\frac{1}{1-i\omega\tau}\,.
\label{eq:response_function}
\eeq
The fluctuation-dissipation theorem \cite{lifshitz1984statistical}
relates the mean-square temperature fluctuation to the imaginary part
of the susceptibility: 
\beq
\langle\delta T^2\rangle = \hbar
\int_{-\infty}^{\infty} \frac{d\omega}{2\pi}\,\,\Im\alpha(\omega)
\coth\frac{\hbar\omega}{2T}.
\label{eq:temp_fluc}
\eeq
The divergence of the integral at high frequencies is fictitious, and
is an artifact of the single-pole form for the response function,
which is just an approximation valid at low frequencies.  
The integral cannot be carried out exactly, but it can easily be
approximated in various limits \cite{balatsky2003quantum}.
In particular the classical result (\ref{eq:classical_fluctuations})
is reproduced at $T\tau\gg \hbar$, by replacing the $\coth$ by the
inverse of its argument. In the opposite regime of $T\tau\ll \hbar$, 
\begin{equation}\label{temp-fluct-quantum}
\langle\delta T^2\rangle\simeq \frac{\hbar T}{\pi C_v\tau }\,\ln\omega _c\tau, 
\end{equation}
where $\omega _c$ is a cutoff frequency, see App.~\ref{app:A} for details. 

Thus far we have been working in the approximation in which the
dominant relaxation is given by an overdamped mode.  In studying
temperature fluctuations of the AdS black hole we will encounter a
different situation, when relaxation to equilibrium is driven by
slowly decaying oscillations. The above discussion then needs to be
modified. To this end, we consider the damped harmonic oscillator with
an internal frequency $\Omega$ and relaxation rate $\Gamma$. The
response function, or the retarded Green function, is given by  
\beq
\alpha(\omega) = G^R(\omega) = -\frac{\Omega^2}{k} \frac{1}{\omega^2 - \Omega^2 + i2\omega \Gamma}.
\eeq
Here $k$ is the ``spring constant'' which in our case is given by
$k=\frac{C_v}{T}$ since the change in free energy is $\Delta S \delta
T = \frac{C_v}{T} \delta T^2$.  So the response function is given by 
\beq
\alpha(\omega) = -\frac{T\Omega^2}{C_v}
\frac{\omega^2-\Omega^2-i2\omega\Gamma}
{(\omega^2-\Omega^2)^2+4\omega^2\Gamma^2}.
\label{eq:BHresponse}
\eeq
The two oscillation poles in the complex frequency plane are located at
\begin{align}
\omega &= -i\Gamma \pm \sqrt{\Omega^2-\Gamma^2} 
\simeq -i\Gamma \pm \Omega. \nonumber
\end{align}
The approximate equality on the right holds for
$\Omega\gg\Gamma$, the case that we will encounter shortly in the
holographic setup. 

Depending on the ratios between the three parameters $T$, $\Omega$
and $\Gamma$, various regimes of temperature fluctuations are
possible. When the temperature is much bigger than both the
frequency and the decay rate, the classical result
(\ref{eq:classical_fluctuations}) is recovered. As may be expected, this result does not depend on a particular form of the response function, and actually follows on very general grounds from the Kramers-Kronig relation
\begin{equation}
 \mathop{\mathrm{Re}}\alpha (\omega ')=-\!\!\!\!\!\!\int_{-\infty }^{+\infty }
 \frac{d\omega }{\pi }\,\,\frac{\mathop{\mathrm{Im}}\alpha (\omega )}{\omega -\omega' }\,.
\end{equation}
 Indeed, setting $\omega '=0$, and taking into account that
\begin{equation}\label{KK-contraint}
 \mathop{\mathrm{Re}}\alpha (0)=\alpha (0)=\frac{T}{C_v}\,,
\end{equation}
we get (\ref{eq:classical_fluctuations}) from (\ref{eq:temp_fluc}) whenever$\coth$ can be replaced by the inverse of its argument. The constraint (\ref{KK-contraint}) will be a useful consistency check on our holographic calculations of the response function.
For the response function of the form (\ref{eq:BHresponse}),
the leading quantum
corrections can be readily computed, here for simplicity displayed in
the oscillatory regime $T/\hbar \gg\Omega\gg\Gamma$ (see App.~\ref{app:A}):
\beq
\langle\delta T^2\rangle \simeq 
\frac{T^2}{C_v}
+ \frac{\hbar^2\Omega^2}{12C_v}
- \frac{\hbar^3\zeta(3)\Omega^2\Gamma}{2\pi^3 C_v T}
+ \mathcal{O}\left(\frac{\hbar^4\Omega^2\Gamma^2}{T^2}\right).
\eeq
The first two terms come from the poles in the response function and
the third term is due to a summation of Matsubara modes. 

The ``quantum'' regime comes about when 
$T/\hbar\ll {\rm max}(\Omega,\Gamma)/2\pi$, which when
$\Omega\gg\Gamma$ yields 
\beq
\langle\delta T^2\rangle \simeq 
\frac{\hbar T\Omega}{2C_v}.
\eeq
A careful analysis is carried out in App.~\ref{app:A} where we define
the classicality parameter:
\beq
\mathfrak{q} \equiv \frac{2\pi T}{\hbar \sqrt{\Omega^2 + \Gamma^2}}.
\label{eq:q}
\eeq
If $\mathfrak{q}>1$ the temperature fluctuations are in the classical
regime, while for $\mathfrak{q}<1$ the temperature fluctuations are
quantum. 

The overdamped regime is recovered when $\Gamma \gg \Omega$. Then a
new scale emerges, namely $\tau=\Gamma/\Omega^2$, which plays the
r\^ole of the relaxation time.  
In this regime the temperature fluctuations obey
\beq
\langle\delta T^2\rangle \simeq \frac{\hbar T}{\pi C_v\tau}
\log\Gamma\tau,
\eeq
which coincides with (\ref{temp-fluct-quantum}) provided that the
cutoff frequency $\omega _c$ is identified with $\Gamma$.  

\section{\label{Sec:2} Local temperature fluctuations of a black hole} 

The system we are interested in is a black hole in the space of
constant negative curvature (AdS${}_{d+2}$), which is dual to a
strongly-coupled CFT in $d+1$ spacetime dimensions, heated to a temperature
that coincides with the Hawking temperature of the black hole. For the most
part, we consider the four-dimensional black-hole ($d=2$), but many
formulas in this section are valid in any number of
dimensions. Since the neutral AdS black hole becomes thermodynamically
unstable at low temperatures,  we  shall consider a larger class of
solutions, namely AdS-Reissner-Nordstr\"om black holes which carry non-zero
charge and are dual to a CFT at non-zero chemical potential. This will
allow us to probe the low-temperature regime when quantum effects are
expected to be important. 

The metric of the AdS${}_4$ black hole under consideration is
\beq
ds^2 = -f(r) dt^2 + \frac{dr^2}{f(r)} 
  + r^2 \left(d\theta^2 + \sin^2 \theta \,d\varphi^2\right),
\label{eq:AdS}
\eeq
where
\beq
f(r) = 1 - \frac{2 M}{r} + \frac{Q^2}{r^2} + \frac{r^2}{R^2}.
\eeq
The temperature, entropy, chemical potential\footnote{This is the
  electrostatic potential of the black hole, identified with the
  chemical potential of the dual field theory by the AdS/CFT
  correspondence.} and extremal charge of the black hole are given by 
\beq
T = \frac{1 + 3\frac{r_+^2}{R^2} - \frac{Q^2}{r_+^2}}{4 \pi r_+}
= \frac{1 + 3\frac{r_+^2}{R^2}}{4\pi r_+}\left(1-\frac{Q^2}{Q_{\rm
   ext}^2}\right), 
\qquad 
S = \pi r_+^2, \qquad 
\mu  = \frac{ Q}{r_+}, \qquad 
Q_{\rm ext} = r_+ \sqrt{1 + 3 \frac{r_+^2}{R^2}},
\label{eq:vals}
\eeq
where  $r_+$ is the horizon radius,
defined as the largest root of $f(r_+)=0$. We absorb the Planck
mass $M_{\rm pl}$ into the definition of the parameters $M$ and $Q$, which
now have the dimension of length. 
The heat capacity of the black hole is given by \cite{Peca:1998cs}
\begin{equation}
C_v = 2\pi r_+^2 \,
\left(\frac{3r_+^2 - R^2}{3r_+^2 + R^2} + \frac{Q^2}{Q_{\rm
    ext}^2}\right)^{-1}
\left(1-\frac{Q^2}{Q_{\rm ext}^2}\right),
\end{equation}
which vanishes, along with the temperature, for the extremal black hole. 

The dual CFT is defined on $S^2\times \mathbb{R}_t$; the sphere has
radius $R$, because far away from the horizon the metric
(\ref{eq:AdS}) asymptotes to 
\beq
ds^2 \simeq  \frac{r^2}{R^2} \,ds_{\rm boundary}^2
+\frac{R^2}{r^2}\,dr^2,\qquad ds_{\rm boundary}^2=-dt^2 + R^2(d\theta^2 + \sin^2 \theta\, d\varphi^2) .
\label{eq:sphere}
\eeq
We will be interested in temperature fluctuations
of the dual field theory on the boundary of spacetime. 
Let $\delta T(\theta ,\varphi)$ be the difference between the local
temperature at the point $(\theta ,\varphi )$ on $S^2$ and the
Hawking temperature.  It is convenient to expand the temperature
difference in spherical harmonics: 
\begin{equation}
 \delta T(\theta ,\varphi )=\sum_{lm}^{}\delta T_{lm}Y_{lm}(\theta ,\varphi ).
\end{equation}
The spherical functions are assumed to be canonically normalized:
\begin{equation}\label{norm-Y}
 \int_{S^d}^{}d^dx\,\sqrt{g}\,Y^*_{lm}(x)Y_{l'm'}(x)=V\delta _{ll'}\delta _{mm'},
\end{equation}
where $V$ is the surface area of the sphere. The two-point correlation
function of temperature fluctuations, by rotational symmetry, should
be independent of the magnetic quantum numbers. We thus define the
power spectrum of temperature fluctuations in the $l$-th harmonics as 
\begin{equation}
 \left\langle \delta T^*_{lm}\delta T_{l'm'}\right\rangle\equiv 
 \left\langle \delta T_l^2\right\rangle\delta _{ll'}\delta _{mm'}.
\end{equation}

At this point it is important to clarify the definition of the local
temperature, the object that will be studied in the rest of the
paper. The inverse Hawking temperature is identified with the
Euclidean-time periodicity of the black hole solution which is a
global quantity. Similarly, in the dual CFT the temperature is not
expressed via any local operators, which renders the treatment of its
local fluctuations ill defined. One finds a way out by considering
the state in which the local thermodynamic equilibrium is
achieved. Indeed, in this case the local thermodynamic equation of
state holds, which expresses the temperature at point $x$ as a function of energy
density $\epsilon$ and density of the number of particles $n$  
\begin{equation}
\label{local}
T(x) = T\big(\epsilon(x), n(x)\big).
\end{equation}
Generally this assumes that thermodynamic quantities do not change
significantly on the scale of the wave length considered in the
problem. Note that now the temperature may be defined locally in the
CFT as it is expressed through the local operators of the
theory. These are the time components of the energy momentum tensor
$T_{00} = \ep$ and the current $J_0 = n$. Furthermore, the
fluctuations of the temperature to linear approximation are given by 
\begin{equation}
\label{oper}
\delta T(x) =  \frac{\p T}{\p \ep}\bigg|_{n} \de T_{00}(x) +  \frac{\p T}{\p n}\bigg|_{\ep} \de J_0(x),
\end{equation}
where $\frac{\p T}{\p \ep}\big|_{n}$ is just the inverse of the
volumetric heat capacity $c_v$. It should be stressed that in the
linear response approximation we substitute the operators, which enter
the coefficients in (\ref{local}), by their expectation values, so
(\ref{oper}) can be considered as a linear combination of the
operators $\de T_{00}(x)$ and $\de J_0(x)$. 
 
The fluctuation-dissipation theorem then expresses the power spectrum
of temperature fluctuations through the retarded correlator of the
local temperature (\ref{oper}): 
\begin{eqnarray}
\label{<dT_l^2>}
 \left\langle \delta T_l^2\right\rangle&=&\frac{1}{V}
 \int_{-\infty }^{+\infty }\frac{d\omega }{2\pi }\,\,
 \mathop{\mathrm{Im}}G^T_l(\omega )\coth\frac{\omega }{2T},
\\
\label{G_l}
 G^T_l(\omega )&=&i\int_{0}^{\infty }dt\,\,{\rm e}\,^{i\omega t}\int_{}^{}d^dx\,\sqrt{g}\,Y_{l0}(x) G^T (t,x),\\
 \label{<TT>}
 G^T (t,x) & =& \frac{\p T}{\p \ep}\bigg|_{n}^{2} \left\langle \left[T_{00}(t,x),T_{00}(0,0)\right]\right\rangle\\
 \notag
 & +& \frac{\p T}{\p \ep}\bigg|_{n} \frac{\p T}{\p n}\bigg|_{\ep} \Big( \left\langle \left[T_{00}(t,x),J_{0}(0,0)\right]\right\rangle + \left\langle \left[J_{0}(t,x),T_{00}(0,0)\right]\right \rangle \Big) \\
 \notag
 & +& \frac{\p T}{\p n}\bigg|_{\ep}^{2} \left\langle \left[J_{0}(t,x),J_{0}(0,0)\right]\right \rangle .
\end{eqnarray}
The factor of $1/V$ in the first equation arises because of the
normalization (\ref{norm-Y}) of the spherical functions. 

\section{Hydrodynamic approximation: neutral case \label{Sec:Hydro}}

We see that the temperature fluctuations generally get contributions 
from both energy density and charge density correlators, as well as
from the mixed ones. We first study the case of the neutral black
hole, i.e.~CFT without a chemical potential, in which the charge
density vanishes and the temperature is directly related to the energy
density  
\begin{equation}
\de T = \frac{1}{c_v} \de T_{0 0}. 
\end{equation}
Hence only the two-point correlator of the energy density remains in
(\ref{<TT>}). 
It can be calculated holographically by studying the response of the
gravitational background to scalar metric perturbations. The retarded
two-point function is then expressed in terms of the quasinormal modes
(QNMs) of the black hole
\cite{Policastro:2002se,Policastro:2002tn}. For a black hole 
with a flat horizon, the lowest QNMs exhibit hydrodynamic behavior
consistent with shear and sound modes of the thermalized plasma state
of the dual CFT \cite{Policastro:2002se,Policastro:2002tn}. The
hydrodynamic approximation should still be accurate for a sufficiently
large black hole with a spherical horizon. Indeed one can calculate
the lowest QNMs of a large AdS black hole from hydrodynamics on
the sphere, without any recourse to Einstein's equations
\cite{Friess:2006kw, Michalogiorgakis:2006jc}. In this section we
compute the response function in the same hydrodynamic approximation,
which is valid in the high-temperature regime, $TR\gg 1$. 

The hydrodynamic equations of motion follow from the conservation 
of the energy-momentum tensor:
\beq
\nabla_\mu T^{\mu \nu} =0,
\eeq
where
\begin{equation}
T^{\mu \nu} = \epsilon u^\mu u^\nu + p \delta^{\mu \nu} - \eta \Delta^{\mu \alpha} \Delta^{\nu \beta} \left(\nabla_\alpha u_\beta + \nabla_\beta u_\alpha - \frac{2}{d} \eta_{\alpha \beta} \nabla_\mu u^\mu \right) - \zeta \Delta^{\mu \nu} \nabla_\lambda u^\lambda.
\end{equation}
Here $u_\mu$ is the local 4-velocity of the liquid (satisfying 
$u^\mu u_\mu =-1$), $\epsilon$ is its energy density, $p$ is the
pressure, $\eta$ and $\zeta$ are the shear and bulk viscosities, and
the covariant derivative, $\nabla_\mu$, is taken with respect to the
boundary metric \eqref{eq:sphere}.  
For a conformal theory, $T_\mu^\mu=0$ leads to $\zeta=0$ and
$\epsilon=dp$. \footnote{One might worry that once a scale is
  introduced in the problem, such as temperature or chemical
  potential, the equality $T_\mu^\mu=0$ no longer holds. This is
  not the case, because the tracelessness of the energy-momentum
  tensor is a feature of the operator algebra of the CFT and thus does 
  not depend on the particular thermodynamic state of the system,
  which may be described by finite temperature and chemical
  potential.} 

The standard way to compute the two-point function of the
energy-momentum tensor is to study a response to metric
perturbations. Then, 
\begin{equation}\label{deltaT}
 \left\langle T^{\mu \nu }(x)
 \vphantom{T^{\mu \nu }(x)T^{\lambda \rho }(y)}
 \right\rangle_{g+h}=\left\langle T^{\mu \nu }(x)
 \vphantom{T^{\mu \nu }(x)T^{\lambda \rho }(y)}
 \right\rangle_{g}+\frac{i}{2}\int_{}^{}d^{d+1}y\,\sqrt{|g|}
 \left\langle T^{\mu \nu }(x)T^{\lambda \rho }(y)\right\rangle_g
 h_{\lambda \rho }(y)+O\left(h^2\right).
\end{equation}
To find the correlation function for the energy density we thus need
to linearize the hydrodynamic equations in the presence of a small
lapse function $h_{00}$ on top of the metric of $S^d\times
\mathbb{R}_t$. The linearized Navier-Stokes equations on $S^d\times
\mathbb{R}_t$ can be found in
\cite{Friess:2006kw,Michalogiorgakis:2006jc}. Keeping track of the
non-zero lapse function, we can recover the source term. This results
in a coupled system of two linear equations: 
\begin{eqnarray}
 \begin{pmatrix}
  \partial _t & \left(\epsilon +p\right) \\ 
  v_s^2\nabla^2 & \left(\epsilon +p\right)\partial _t-\frac{2}{d}\,\eta \mathcal{R} -\left(\zeta +2\,\frac{d-1}{d}\,\eta \right)\nabla^2\\ 
 \end{pmatrix}
 \begin{pmatrix}
  \delta \epsilon  \\ 
  \nabla_i\delta u^i \\ 
 \end{pmatrix}=
 \begin{pmatrix}
  0 \\ 
  \frac{1}{2}\left(\epsilon +p\right)\nabla^2h_{00} \\ 
 \end{pmatrix},
\end{eqnarray}
where $\mathcal{R}=d(d-1)/R^2$ is the Ricci curvature and $-\nabla^2$
is the invariant Laplacian on the sphere, and $v_s^2=\partial
p/\partial \epsilon$ is the speed of sound. Expanding in spherical
harmonics, solving for $\delta \epsilon$ and comparing with
(\ref{deltaT}), we get for the Green's function defined in
(\ref{G_l}): 
\begin{equation}\label{lthmode}
 G_l(\omega )=-\frac{\epsilon +p}{c_v^2 v_s^2}\,\,\frac{\Omega^2_l}{\omega^2 - \Omega^2 _l+2i\omega \Gamma_l}
\end{equation}
with
\begin{eqnarray}\label{Omega-l}
 \Omega _l&=&\frac{v_s}{R}\,\sqrt{l\left(l+d-1\right)}
 \\
 \label{Gamma-l}
 \Gamma_l &=&\frac{1}{\left(\epsilon +p\right)R^2}\left[\frac{\left(d-1\right)\left(l+d\right)\left(l-1\right)}{d}\,\eta +\frac{l\left(l+d-1\right)}{2}\,\zeta \right].
\end{eqnarray}
The correct normalization of the response function, as in
(\ref{eq:BHresponse}), follows from (\ref{<dT_l^2>}), (\ref{lthmode})
by virtue of a thermodynamic identity at zero chemical potential 
$$
c_v v_s^2=\frac{\partial \epsilon }{\partial T}\,\,\frac{\partial p}{\partial \epsilon }=\frac{\partial p}{\partial T}=s=\frac{\epsilon +p}{T}.
$$
This guarantees matching to classical thermodynamics
(\ref{eq:classical_fluctuations}) in the high-temperature limit.

Taking into account that for a conformal fluid, $v_s^2=1/d$ and $\zeta
=0$, and using the universal holographic result
\cite{Kovtun:2004de,Policastro:2002se} for the viscosity-to-entropy
ratio, $\eta /s=1/4\pi$, gives for the hydrodynamic QNMs
\cite{Friess:2006kw,Michalogiorgakis:2006jc}: 
\begin{equation}
\omega_{\rm hyd} = \pm \frac{1}{R}\sqrt{\frac{l\left(l+d-1\right)}{d}} 
- \frac{i(d-1)(l+d)(l-1)}{4\pi dTR^2}\,.
\end{equation}
We have taken into account here that $\Omega _l/\Gamma _l\sim RT/l\gg
1$ (unless $l$ is very big, but the hydrodynamic approximation is not
applicable to such high-frequency modes anyway), and so the sound
modes attenuate very slowly. It is also true that $T/\Omega _l\sim
TR/l\gg 1$, which implies that temperature fluctuations are purely
classical as long as hydrodynamics is an accurate approximation. 

It is important to note here, that there are no quasinormal modes with
$l=0$ and $l=1$ in the spectrum. The former would correspond to the
homogeneous change of the energy density in the whole volume of the
system and thus this fluctuation would violate the energy
conservation law. From the point of view of linear response theory
this means that once the system is perturbed by a force, which
homogeneously changes the energy density, it acquires a new equilibrium
state and does not relax to the initial one. The mode with $l=1$
corresponds to a simple rotation in $S^2$. In the case of flat space
it would correspond to a translation along a given direction. Thus in
homogeneous space it is the Goldstone mode, associated with 
translation symmetry. Indeed the shift of the center of mass of the
system will not change its state and hence not produce any
counter-force, so 
there is no relaxation associated with this mode. The argument above
indicates that we are dealing with  \textit{local} temperature
fluctuations. The 
fluctuations of the total temperature may be described only by the
$l=0$ mode, because all the others average to zero upon integration
over the volume of the system. But at the same time for the closed
system with conserved energy, such as the large AdS-RN black hole or the CFT
on a sphere, the fluctuations of the total temperature are forbidden
by the energy conservation law and thus the $l=0$ mode is
absent. Nonetheless, \textit{local} fluctuations of temperature
are allowed and their study is a well-defined problem.

\section{Hydrodynamic approximation: charged case \label{Sec:charge}}

Our discussion so far applies to neutral black holes. The main
complication that arises for charged black holes is that the response
function for temperature relaxation then includes, in addition to
$\left\langle T_{00}T_{00}\right\rangle$, also a contribution from
$\left\langle T_{00}J_0\right\rangle$ and $\left\langle
J_{0}J_0\right\rangle$ as shown in (\ref{<TT>}). To calculate these
additional correlators in the hydrodynamic approach we need to include the
current conservation law into the system of hydrodynamic equations and
consider the external field, coupled to the current, in order to
define Green functions in the variational approach
\cite{Kovtun:2012rj}. Thus we have 
\begin{equation}
\nabla_{\mu} T^{\mu \nu} = F^{\nu \lambda} J_\lambda \qquad \nabla_{\mu} J^{\mu} = 0, 
\end{equation}
where $F^{\mu \nu} = \p_\mu A_\nu - \p_\nu A_\mu$ is an external field, the constitutive equation for the current is
\begin{equation}
J^\mu = n u^\mu - \sigma T \Delta^{\mu \nu} \nabla_\nu (\mu/T) + \sigma \Delta^{\mu \lambda} F_{\lambda \nu} u^\nu 
\end{equation}
with $\sigma$ being the conductivity. In what follows, we will consider
only the scalar potential of the external field $A_0$ which
couples to the charge density operator $n = J_0$. The system of
hydrodynamic equations may now be written as 
\begin{eqnarray}
 \begin{pmatrix}
  \partial _t & w & 0 \\ 
  \beta_1 \nabla^2 & w \partial _t - \frac{2}{d}\,\eta \mathcal{R} - \left(\zeta +2\,\frac{d-1}{d}\,\eta \right)\nabla^2 & \beta_2 \nabla^2 \\
  - \sigma \alpha_1 \nabla^2 & n & \p_t - \sigma \alpha_2 \nabla^2
 \end{pmatrix}
 \begin{pmatrix}
  \delta \epsilon  \\ 
  \nabla_i\delta u^i \\ 
  \de n
 \end{pmatrix}=
 \begin{pmatrix}
  0 \\ 
  \frac{1}{2} w \nabla^2h_{00} - n \nabla^2 A_0 \\
  \sigma \nabla^2 A_0
 \end{pmatrix},
\end{eqnarray}
where $w = \epsilon + p$ is an enthalpy density and we introduced the notation for thermodynamic derivatives
\begin{align}
\label{thermo_der}
\alpha_1 &= T \left. \frac{\p(\mu/T)}{\p \epsilon} \right|_n,
& \alpha_2 &= T \left. \frac{\p(\mu/T)}{\p n} \right|_\epsilon,
\\
 \beta_1 &= \left. \frac{\p p}{\p \epsilon} \right|_{n}, 
& \beta_2 &= \left. \frac{\p p}{\p n} \right|_{\epsilon}. \label{thermo_der2}
\end{align}
The Green function is now a matrix and defines the response of $(\de \eps, \de n)$ on the sources $(h_{00}, \phi)$ as
\begin{equation}
 \Bigg(\begin{matrix} 
  \delta \epsilon \\ \delta n 
 \end{matrix} \Bigg) =  G \  
  \Bigg( \begin{matrix} 
  - \frac{1}{2} h_{00} \\  A_0
 \end{matrix} \Bigg), \qquad 
 G =  \begin{pmatrix} 
  \la T_{00} T_{00} \ra &&   \la J_0   T_{00} \ra \\
  \la T_{00} J_0 \ra && \la J_0 J_0 \ra
 \end{pmatrix}.
\end{equation}
It may be expressed as
\begin{align}
\label{Green}
G = \frac{1}{\mathcal{P}} \begin{pmatrix}
                 k_l^2 w \omega + i \sigma k_l^4 w \alpha_2 &&  k_l^2 n \omega + i \sigma k_l^4 (n \alpha_2 - \beta_2) \\
                 k_l^2 n \omega - i \sigma k_l^4 w \alpha_1 && - i \sigma k_l^2 \omega^2 + k_l^2 \omega \left(\frac{n^2}{w} + 2 \sigma \Gamma_l \right) - i \sigma k_l^4 (n \alpha_1 - \beta_1), 
                \end{pmatrix}
\end{align}
where we adopted a short-hand notation for the spherical ``momentum'' 
$$
k_l^2 = \frac{l (l+d-1)}{R^2}
$$
and $\Gamma_l$ coincides with (\ref{Gamma-l}). The pole structure is governed by the denominator $\mathcal{P}$ and includes two sound modes and one purely imaginary diffusive mode
\begin{equation*}
\mathcal{P} = (\omega^2 - \Omega_l^2 + 2 i \omega \hat{\Gamma}_l) (\omega + i \mathcal{D} k_l^2), 
\end{equation*}
where
\begin{align}
\label{Omega-sound}
\Omega_l^2 &= k_l^2 \left[\beta_1 + \frac{n}{ w} \beta_2 \right] = k_l^2 v_s^2, \\
 \hat{\Gamma} &= \Gamma_l + \frac{1}{2} \frac{k_l^2}{w} \frac{\sigma}{ v_s^2} \beta_2^2, \\
\label{diffusion}
\mathcal{D} &=  \frac{\sigma}{ v_s^2} \big(\alpha_2 \beta_1 - \alpha_1 \beta_2 \big). 
\end{align}
Using the thermodynamic relations which we obtain in Appendix
\ref{Thermo}, one can check several important features of this
result. First of all (\ref{off-diag}) shows that the off-diagonal
terms in the Green function (\ref{Green}) are equal, which is also 
demanded by the Onsager relation. Then according to
(\ref{speed-of-sound}), the speed of sound introduced in
(\ref{Omega-sound}) coincides indeed with the usual result $v_s^2 =
\frac{\de p}{\de \ep} \big|_{s}$.  

Finally, we can check whether the normalization of the Green function
leads to the classical result (\ref{eq:classical_fluctuations}). For
this purpose we plug the expressions for the two-point correlators
into the fluctuation-dissipation theorem (\ref{<dT_l^2>}). Then as
discussed earlier, in order to get the mean amplitude of the 
temperature fluctuations in the classical limit, one considers the
value of the real part of the response function at zero frequency. The
result is 
\begin{align}
\label{dT_classical}
 \la \delta T^2 \ra_{T \rar \infty} =  T \mathop{\rm Re} \la \delta T_l \delta T_l \ra_{\omega \rar 0} = T \frac{\left( \frac{\p T}{\p \epsilon} \big|_{n} \right)^2 \alpha_2 - 2  \frac{\p T}{\p \epsilon} \big|_{n} \frac{\p T}{\p n} \big|_{\epsilon} \alpha_1 + \left( \frac{\p T}{\p n} \big|_{\epsilon}  \right)^2 \alpha_3}{\alpha_1^2 - \alpha_2 \alpha_3} = \frac{T^2}{c_v},
\end{align}
where $\alpha_3 = \frac{1}{T} \frac{\de T}{\de \ep} \big|_{n}$ and we
used the thermodynamic relations (\ref{temperature_numerator}) and
(\ref{alphas}).  

The above considerations are completely general and hold for any kind
of fluid with conserved charge. When considering a CFT, the
thermodynamic expressions simplify considerably. As in the CFT, the
energy is proportional to the pressure $\ep = d \ p$, we get $\beta_2 =
0$, $\beta_1 = v_s^2 = \frac{1}{d}$ and $w \alpha_1 + n \alpha_2 = 0$
(see Appendix \ref{Thermo}). The two-point functions of the operators
under consideration can be expressed as 
\begin{align}
\la \de \ep_l \de \ep_l \ra &=  \frac{w}{v_s^2} \frac{\Omega_l^2}{\omega^2 - \Omega_l^2 + 2 i \Gamma_l \omega}, \\
\la \de \ep_l \de n_l \ra & =  \frac{n}{v_s^2} \frac{\Omega_l^2}{\omega^2 - \Omega_l^2 + 2 i \Gamma_l \omega}, \\
\la \de n_l \de n_l \ra & = \frac{n^2}{v_s^2 w} \frac{\Omega_l^2}{\omega^2 - \Omega_l^2 + 2 i \Gamma_l \omega} + \frac{d n^2 - w \chi }{w} \frac{ i \mathcal{D} k_l^2}{\omega + i \mathcal{D} k_l^2},
\end{align}
where we introduced the susceptibility
\begin{equation}
\chi = \frac{\p n}{\p \mu} \bigg|_{T}.
\end{equation}
Remarkably, the diffusive pole cancels out in the correlators of the
energy density and the position of the sound pole is defined by the
same expression as in the case of the neutral liquid
(\ref{Omega-l}). The expression for the correlator of temperature
fluctuations contains the sum of the contributions from the sound and
the diffusive modes  
\begin{equation}
\label{dT_CFT}
\la \de T_l \de T_l \ra =
 \frac{T^2}{d w}  \frac{\Omega^2}{\omega^2 - \Omega^2 + 2 i \Gamma_l \omega}  + \frac{n^2 T^2}{w (d n^2 -w \chi)}  \frac{i \mathcal{D} k_l^2}{\omega + i \mathcal{D} k_l^2}.
\end{equation}
The classical limit of the temperature fluctuations is of course
reproduced as in the general case (\ref{dT_classical}). 

Interestingly, mixing with  charge-density fluctuations induces a diffusive pole in the response function of temperature fluctuations, which was not there at zero chemical potential.\footnote{Note that the contribution of the diffusive pole in (\ref{dT_CFT}) is of order $n^2$ justifying the analysis of the previous section.} The origin of this pole is easy to understand. Temperature gradients in the density wave create induced currents via the thermoelectric effect. Dissipation of these currents efficiently attenuates temperature fluctuations even at zero frequency.

The quantum
limit may be analyzed separately for the two contributions along the
lines sketched in Section \ref{Sec:1}. At low temperatures the sound
pole gives a contribution 
\begin{equation}
\label{sound}
\la \delta T^2 \ra_{\rm sound} \simeq \hbar \frac{T \Omega}{2} \frac{T}{2 w} =  \hbar\frac{T^2}{4 w}  \sqrt{ \frac{l (l + d -1)}{d}}\frac{1}{R},
\end{equation}
while the diffusive contribution is
\begin{equation}
\label{diffuse}
\la \delta T^2 \ra_{\rm diffusive} \simeq \frac{\hbar}{2 \pi} \frac{T^2 n^2 \mathcal{D}}{w (d n^2 - w \chi)}  \mathrm{log} \left(\frac{\omega_c}{\mathcal{D} k_l^2} \right) \frac{l (l + d -1)}{d} \frac{1}{R^2}.
\end{equation}
One can see that generally in the hydrodynamic approximation the
diffusive contribution is suppressed compared to that from the sound,
because it scales as $R^{-2}$ and $R$ is assumed to be large. More
specifically, the expression (\ref{diffuse}) contains the diffusive
constant $\mathcal{D}$ which by dimensional analysis is inversely
proportional to the temperature. For a neutral CFT, the universal
value for $\mathcal{D}$ was studied in \cite{Kovtun:2008kx}:
$\mathcal{D} \sim \frac{1}{4 \pi T}$. Thus the relation between sound
and diffusive contributions is  
\begin{equation}
\label{SoverD}
 \frac{\la \delta T^2 \ra_{\rm sound}}{\la \delta T^2 \ra_{\rm diffusive}} \sim T R.
\end{equation}
One should be careful though using this estimate to make any
conclusions about the dominance of sound or diffusive mode in the
temperature fluctuations at low temperatures. It was derived in the
hydrodynamic approximation which is, strictly speaking, valid only at
large temperatures $T R \gg 1$, where the quantum limit is not
applicable. Indeed, the transition between classical
(\ref{eq:classical_fluctuations}) and quantum
(\ref{sound}),(\ref{diffuse}) behavior of the fluctuations happens
when the temperature becomes comparable with the characteristic
frequency of the quasinormal mode (see Appendix \ref{app:A}). For the
sound mode it happens when $T \ll \Omega$, i.e. $T R \ll 1$, while for
the diffusive one, $T\ll \mathcal{D} k_l^2$, which means $T^2 R^2 \ll
1$. Hence, we can conclude that the hydrodynamic approach is
incompatible with studying the quantum regime and in order to study the
classical/quantum transition one needs to go beyond hydrodynamics. For
this we will turn to the direct gravitational study of spherical black
holes in the next section, for which the analysis necessitates
numerical calculations.

\section{\label{Sec:3} Quasinormal modes of the spherical black hole}

The quasinormal modes of a spherical AdS-Reissner-Nordstr\"{o}m black hole
were thoroughly studied in 
\cite{Chan:1996yk,Horowitz:1999jd,Cardoso:2001bb,Moss:2001ga,Berti:2003ud} (see \cite{Berti:2009kk} for a review). 
The entire spectrum of quasinormal modes of the Reissner-Nordstr\"om
black hole in AdS${}_4$ was calculated in \cite{Berti:2003ud}, following the
approach developed in \cite{Cardoso:2001bb,Moss:2001ga,Mellor:1989ac}. Although in the
axial gravitational channel an ``exceptional''
frequency was found, which can be related to the hydrodynamic
shear mode \cite{Policastro:2002se}, no long-lived modes were observed in
the polar gravitational channel, which would correspond to the
sound mode \cite{Policastro:2002tn} discussed above. 
This discrepancy with the hydrodynamic results was pointed out in
\cite{Friess:2006kw,Michalogiorgakis:2006jc}. It turned out, that one
should pay special attention to the boundary conditions of the polar
gravitational mode, because simple Dirichlet boundary conditions on
the master field lead to metric fluctuations, which perturb the
asymptotic behavior of AdS${}_4$ spacetime. Instead, special Robin
boundary conditions were found, which do not perturb the asymptotic
metric and lead to quasinormal modes consistent with the
hydrodynamic picture. 
The corresponding numerical calculations were done for the 5- and
4-dimensional AdS-Schwarzschild metric in \cite{Friess:2006kw,Michalogiorgakis:2006jc} and analytic results for long-lived
modes of neutral black holes in AdS-space of any dimension were obtained in
\cite{Siopsis:2007wn}. The analogous treatment of Kerr-AdS black holes was
made in \cite{Cardoso:2013pza,Dias:2013sdc}.
The study of quasinormal frequencies of the sound mode in
an AdS-Reissner-Nordstr\"om black hole with the aforementioned Robin
boundary conditions has, to the best of our knowledge, not been done
before.\footnote{We note, that the spectrum of AdS${}_{5-7}$ 
  Reissner-Nordstr\"om black holes was obtained in
  \cite{Konoplya:2008rq}, but the Dirichlet boundary conditions were
  used for the master function at the AdS boundary and quasinormal
  modes calculated in this way may not coincide with the hydrodynamic
  ones.}  

In what follows we focus on the calculation of the quasinormal modes
of the charged black hole in AdS${}_4$, i.e.~we choose the number of
spatial directions in the dual CFT to be $d=2$.
In order to proceed we first need to figure out, which particular
boundary conditions one should use in this case in order to obtain
results consistent with hydrodynamics. The details of our treatment
can be found in Appendix \ref{app:B}.

There are two families of quasinormal modes of the charged black hole
which are found as eigenmodes of the Sturm-Liouville problem defined
by the master equations
\cite{chandrasekhar1998mathematical,Mellor:1989ac} 
\begin{equation}
\label{Zs}
\frac{\Delta}{r^2} \frac{d}{dr}\left(\frac{\Delta}{r^2} \frac{d}{dr} Z_i^+ \right) + \omega^2 Z_i^+ = V_i^+ Z_i^+ \qquad (i=1,2) 
\end{equation}
where $\Delta = r^2 - 2 Mr + Q^2 + \frac{r^4}{R^2}$ and the
Schr\"odinger-type potentials are given in (\ref{potentials}). In the zero
charge limit the mode $Z_1^+$ describes purely electromagnetic
excitations, while $Z_2^+$ reduces to a purely gravitational ones. At
non-zero charge the modes get mixed similarly to the hydrodynamics case,
but it is still useful to denote them as ``mostly-electromagnetic''
and ``mostly-gravitational''. According to the AdS/CFT duality the
gravitational excitations of black hole are dual to the fluctuations
of energy-momentum tensor $\de T_{\mu \nu}$ of the dual CFT and the
electromagnetic excitations are dual to the current $\de J_\mu$. Thus
we expect that the quasinormal frequencies of $Z_2^+$ will match the
hydrodynamic sound mode when the radius of the black hole horizon is
large and similarly $Z_1^+$ will have a diffusion-type QNM.

The important part of the formulation of the Sturm-Liouville problem
is the boundary conditions. In Appendix \ref{app:B} we find that in
order to avoid perturbing the metric at the asymptotic AdS boundary
one needs to impose Robin-type boundary conditions on $Z_i^+$ as was
suggested in \cite{Friess:2006kw, Michalogiorgakis:2006jc} for the
case of the neutral black hole. For the charged case we get (see
eq.~\ref{bcB}) 
\begin{align}
 r^2 \frac{\p_r Z^+_1}{Z^+_1}\bigg|_{r \rar \infty} &= \left(\frac{3 M}{n} - \frac{p_1}{2 n} \right), \\
 r^2 \frac{\p_r Z^+_2}{Z^+_2}\bigg|_{r \rar \infty} &= \left(\frac{3 M}{n} + \frac{4 Q^2 }{p_1} \right),
\end{align}
where $p_1 = 3M + \sqrt{9 M^2 + 4 Q^2 (l-1)(l+2)}$. On the horizon we
adopt the purely infalling wave boundary conditions, which are easily
expressed in ``tortoise'' coordinates; $dr^* = \frac{r^2}{\Delta} dr$,
\begin{equation}
Z_i^+ \Big|_{r_* \rar -\infty} \sim e^{-i \omega(\tau  + r_*)}.
\end{equation}

\begin{figure}[!ht]
\begin{tabular}{cc}
\includegraphics[width=0.45 \linewidth]{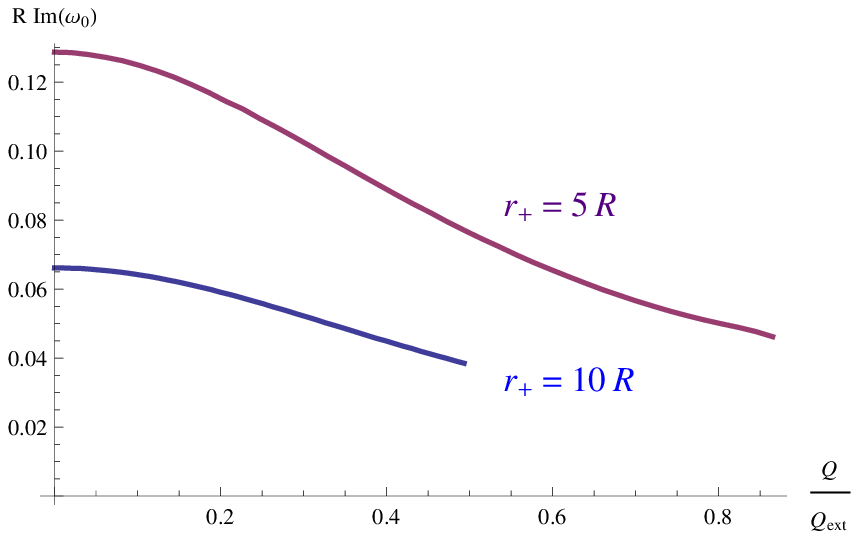} & 
\includegraphics[width=0.45 \linewidth]{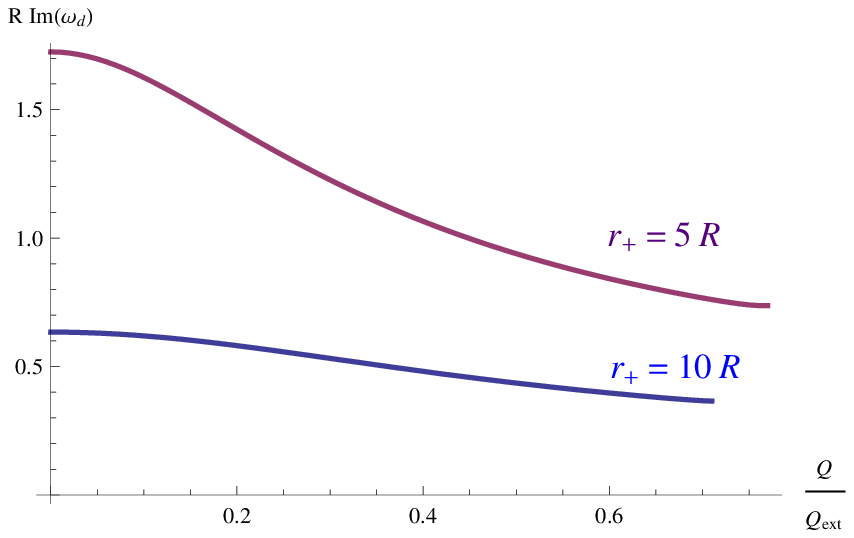} \\
a) & b)
\end{tabular}
\centering
\caption{\label{plot_QNM} Imaginary part of the quasinormal frequency
  of the gravitational mode with $l=2$ for $r_+=5R$ and $r_+=10R$ as a
  function of the black hole charge: a) the ``mostly-gravitational'' or hydrodynamic sound mode; b) the ``mostly-electromagnetic'' or hydrodynamic diffusive mode. The frequency is 
  given in units of the AdS curvature-radius $R$.} 
\end{figure}

We study the quasinormal frequencies $\omega_0$ of the gravitational
polar mode (related to the dual hydrodynamic sound mode) for 
different choices of charge and horizon radii of the black hole. The
resulting frequencies at $r_+=5 R$ and $r_+ = 10 R$ for
angular momentum $l=2$ are shown in Fig.~\ref{plot_QNM}a). The curves 
demonstrate similar behavior to that observed for the
``exceptional'' mode in the axial channel in \cite{Berti:2003ud}. 
For a neutral black hole our results coincide with those presented in
\cite{Michalogiorgakis:2006jc} for $r_+ \lesssim 20 R$, which is the range
over which our numerical calculation yields reliable precision. The
behavior of the quasinormal frequencies of the electromagnetic mode is
shown in Fig.~\ref{plot_QNM}b). The quasinormal frequency for this
mode is purely imaginary as expected for the diffusive mode and
behaves similar to the imaginary part of the sound mode, but is
almost 10 times larger. 

At finite electrostatic potential it is especially interesting to
compare the results of our computation to the hydrodynamic
approximation. As it was shown in Section \ref{Sec:charge}, the position of the sound mode in the charged and neutral liquid is given by the same expression (\ref{Omega-l}--\ref{Gamma-l}). Using the thermodynamic relation $\epsilon + p = T s + \mu\rho$ and universal value for the shear viscosity $\eta = s/4 \pi$ \cite{Kovtun:2004de}, 
we get: 
\begin{equation}
\omega_{\rm hyd} = \pm \frac{1}{R}\sqrt{\frac{l(l+d-1)}{d}} 
- \frac{i(d-1)(l+d)l(l-1)}{4\pi d R^2}\frac{1}{T+\mu\frac{Q}{S}},
\notag
\end{equation}
where $S$ and $Q$ are the total entropy and charge of the system. 
Substituting the values for temperature, charge, potential
and entropy of the dual Reissner-Nordstr\"om black hole \eqref{eq:vals}, we
obtain the following final result for the sound mode frequency in the
hydrodynamic regime
\begin{equation}
\omega_{\rm hyd} = \pm \frac{1}{R}\sqrt{\frac{l(l+d-1)}{d}} 
- \frac{i(d-1)(l+d)(l-1)}{d}\frac{r_+}{R^2+3r_+^2+3\mu^2R^2}.
\label{eq:hydro}
\end{equation}

\begin{figure}[!ht]
\includegraphics[width=0.49 \linewidth]{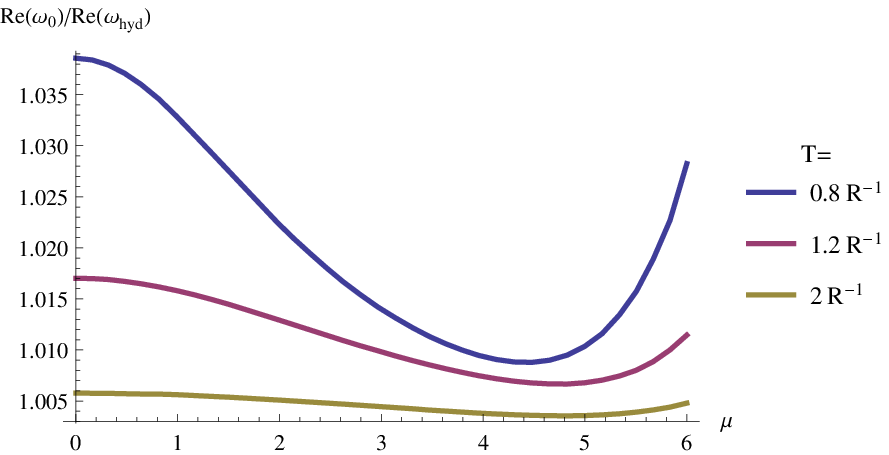}
\includegraphics[width=0.49 \linewidth]{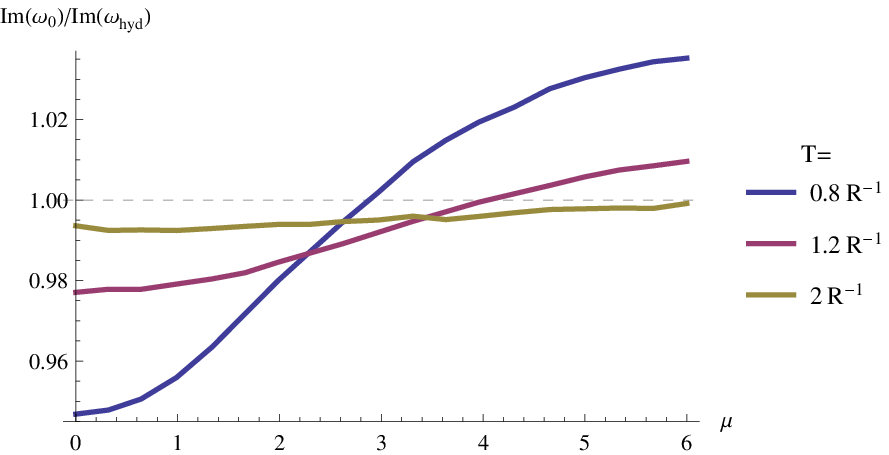}
\centering
\caption{\label{TPhiFig} The relative values of the real and imaginary
  part of 
  the quasinormal frequency at $l=2$ with respect to that given in the
  hydrodynamic approximation \eqref{eq:hydro}.} 
\end{figure}

Fig.~\ref{TPhiFig} shows the
relation between the numerical gravitational result and the hydrodynamic one for $l=2$ mode for
different black hole temperatures. One can see that the curves approach unity
quite fast and already at $T \approx 2 R^{-1}$ the discrepancy 
is less than one percent, which is comparable to our numerical
precision. This is a valuable check of the applicability of our
procedure for calculating quasinormal modes. 

\begin{figure}[!ht]
\includegraphics[width=0.49 \linewidth]{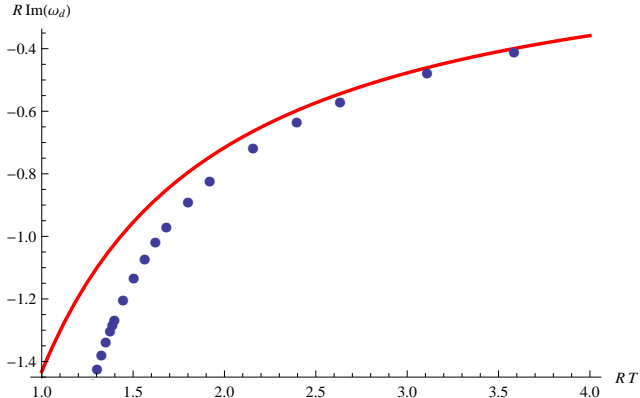}
\centering
\caption{\label{compare_hydro_EM} Comparison of the quasinormal
  frequencies in the diffusive/electromagnetic channel at zero
  chemical potential. Blue dots are the numerical results for QNMs of
  $Z_1^+$ for a black hole (\ref{Zs}) while the red line is the
  hydrodynamics result for the CFT in $d=2$ \eqref{Dhyd}.}  
\end{figure}

In case of the electromagnetic/diffusive channel we can perform a
similar comparison with hydrodynamics. In \cite{Kovtun:2008kx} the
diffusive constant for the CFT with zero chemical potential was
obtained for any number of spatial directions\footnote{We should
  point out that the definition of $d$ in our paper is different from
  that of \cite{Kovtun:2008kx}. We use $d$ as the number of spatial
  dimensions of the CFT, which is equal to 2 in the case under
  consideration. In \cite{Kovtun:2008kx} $d$ is instead the number of
  spacetime dimensions. }  
\begin{equation}
\label{Dhyd}
\mathcal{D}_{hyd}\big|_{\mu=0} = \frac{d+1}{d-1} \frac{1}{4 \pi T}.
\end{equation}
The comparison of our numerical results for the electromagnetic modes
with this expectation is shown in Fig.~\ref{compare_hydro_EM}. Again
one can see, that the hydrodynamic approximation works quite well at
relatively small temperature $T \sim 3 R^{-1}$. 

\begin{figure}[hp]
\includegraphics[width=0.9 \linewidth]{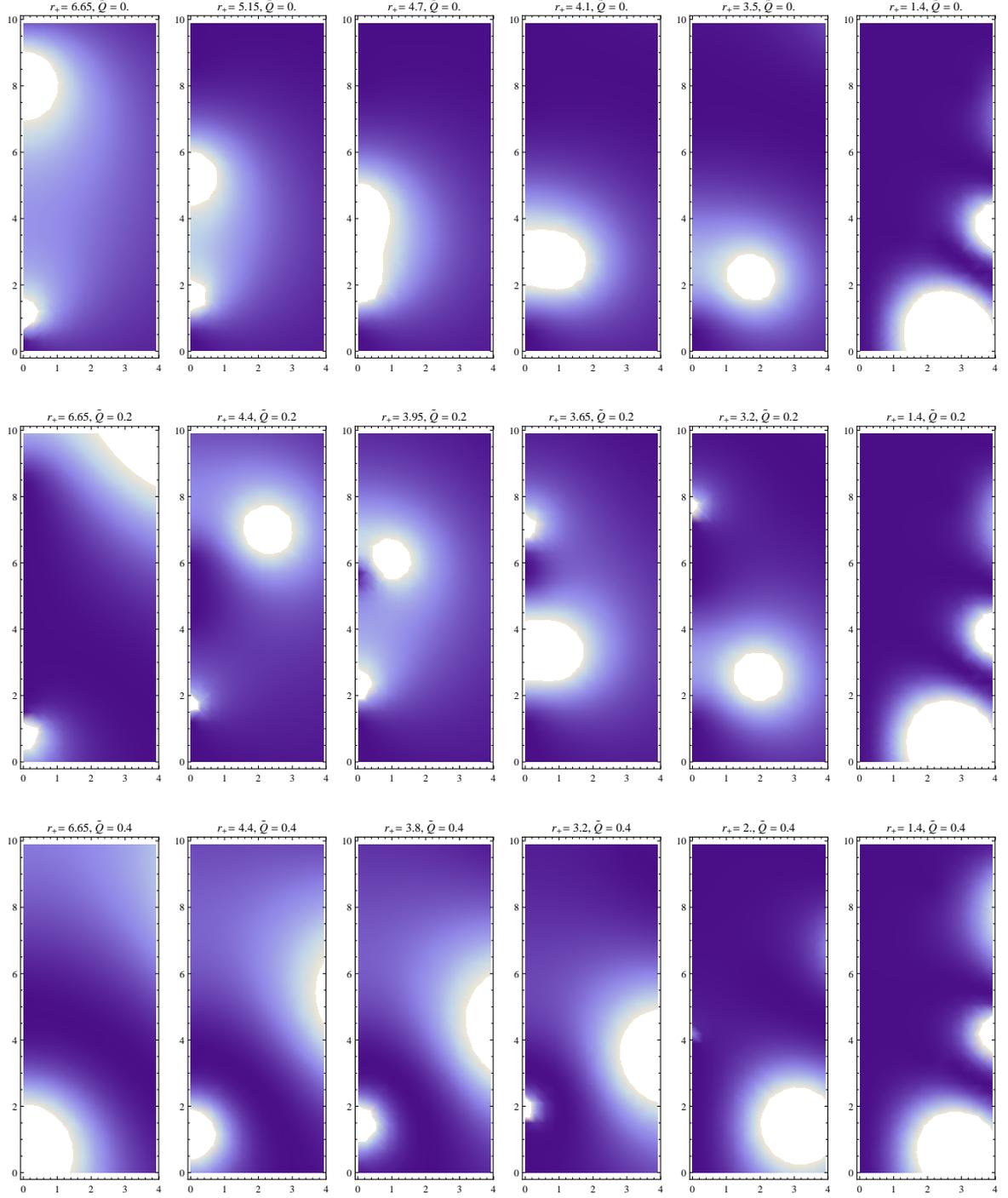}
\centering
\caption{\label{EM_poles_grid} Position of the QNMs in the
  ``mostly-electromagnetic'' channel of the Reissner-Nordstr\"om black
  hole for various radii and charges in the complex $\omega$
  plane. $\tilde{Q} = Q/Q_{ext}$, $r_+$ is measured in units of $R$
  and frequency is measured in units of $R^{-1}$. } 
\end{figure}


Using the numerical values of the quasinormal frequencies for various
$T$ and $\mu$, we can calculate the value of the parameter
$\mathfrak{q}$, which measures the classicality of the temperature
fluctuations \eqref{eq:q}. It is interesting to study the transition
to the quantum regime in both sound and diffusive contributions in
(\ref{dT_CFT}). Looking at the behavior of the quasinormal frequencies
in Fig.~\ref{plot_QNM} one might conclude, that the diffusive mode
violates the bound $\mathfrak{q} > 1$ at large temperatures, as its
quasinormal frequency is numerically large and, moreover, is inversely
proportional to the temperature (\ref{Dhyd}). However, observing the
behavior of the actual gravitational quasinormal modes we see an
interesting feature depicted on Fig.~\ref{EM_poles_grid}. Starting
from large radii of the black hole, we see that in case of the
neutral black hole the imaginary part of the hydrodynamic eigenmode
first grows when the temperature is decreased, as expected from
(\ref{Dhyd}), but reaching the values $\approx i 2 R^{-1}$ it collides
with another purely imaginary mode and recombines developing a pair
of oscillating modes, which then travel back to the real axis
maintaining the absolute value $|\omega_d|\approx 2 R^{-1}$. This
second purely imaginary mode behaves as $\omega \sim r_+$ and is a
member of a series of the QNMs that one can typically find in the
spectrum of spherical black holes \cite{Cardoso:2001bb,
  Berti:2003ud,Berti:2009kk}. 

Turning on a charge of the black hole does not change the picture
substantially; again the hydrodynamic mode first moves away from the
origin, but then it gets interfered with and substituted by the lowest
mode from the series of spherical QNMs. Importantly, the mode which is
closest to the origin never has an absolute value which is
significantly bigger than $2 R^{-1}$. Interestingly, the recombination
of poles happens exactly in the region, where $\mathfrak{q} \approx 1$
for the diffusive mode. 

A similar picture is found in the spectrum of
the sound mode. The real part of the
mode is defined by the speed of sound and does not change
significantly with temperature, the imaginary part is numerically
suppressed by a factor of 10, so its growth is negligible. At small
radii the series of spherical modes with $|\omega| \sim r_+$ moves
close to the hydrodynamic one. Again 
the closest mode is never farther from origin than $2 R^{-1}$. 
These considerations allow us to draw the following conclusion. The
transition to the quantum regime at low temperatures in the quantity
$\mathfrak{q} = \frac{2 \pi T}{|\omega_0|}$ happens for both modes
approximately simultaneously, as the absolute values of the lowest
modes $|\omega_0|$ are frozen near $2 R^{-1}$ while the temperature
keeps decreasing.

\begin{figure}[!ht]
\includegraphics[width=0.5 \linewidth]{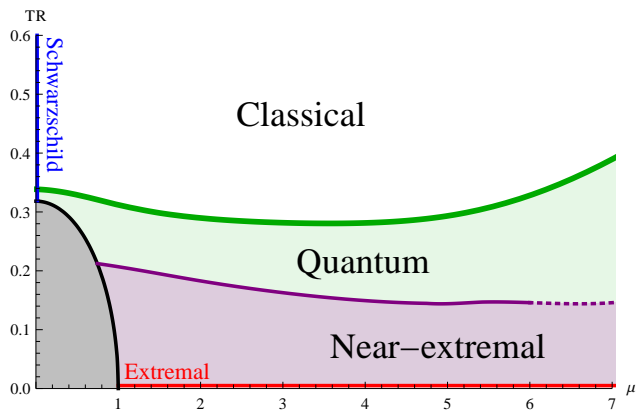}
\centering
\caption{\label{phases} Phase structure of the charged black hole in
  AdS${}_4$ in the grand canonical ensemble. The black line marks the
  ``Hawking-Page type'' phase transition \cite{Chamblin:1999tk}. The green
  line is $\mathfrak{q} = 1$, and thus in the green region quantum
  effects become important. The purple line marks the region, where
  the numerically computed fluctuation $\left\langle \delta T^2
  \right\rangle^{1/2}$ is comparable to the 
  temperature itself (at $\mu >6$ our precision does not allow us to
  continue the line reliably).}  
\end{figure}

To illustrate this transition we plot the curve where 
$\mathfrak{q}=1$ -- for simplicity for the sound mode -- in the phase
diagram of the charged black hole in AdS space in the grand canonical
ensemble, i.e.~for fixed temperature $T$ and asymptotic electrostatic 
potential~$\mu$; see Fig.~\ref{phases}. It is not surprising, that at
a decent value of the potential the curve behaves almost like a
constant temperature line. 

The phase diagram in Fig.~\ref{phases} contains also the phase
transition between the Reissner-Nordstr\"om black hole and the
thermal gas of particles in the background of the extremal black hole
with given potential $\mu$, denoted by AdS${}^{*}$ \cite{Chamblin:1999tk}. At
zero charge this phase transition reduces to the Hawking-Page
transition for the AdS-Schwarzschild black hole \cite{Hawking:1982dh}. We
find that the lowest value of the parameter $\mathfrak{q}$, which is
achieved before this phase transition occurs, is 
$\mathfrak{q} \approx 0.92$ so the deep quantum regime is never reached
for the 
neutral black hole. This is also supported by the peculiar behavior of the diffusive mode, discussed above. Instead one should speak about the region where
quantum effects are becoming important. 

The other interesting line in the phase diagram is the one describing
the region, where the temperature fluctuations computed numerically via
(\ref{eq:temp_fluc}) using (\ref{eq:BHresponse}) for the gravitational  
quasinormal modes are of the order of the temperature itself. 
Linear response theory is no longer reliable below this line and it 
becomes meaningless to speak of small fluctuations, so we can plot it
only approximately. 
Extremal black holes are thus not included in our treatment. 

\section{\label{Conclusion} Discussion}

We have studied temperature fluctuations in
the holographic dual of the AdS-Reissner-Nordstr\"om black
hole. Interestingly, temperature fluctuations are described by
quasinormal modes of a black hole in both the sound channel, for which
the oscillation frequencies are much larger than the attenuation
rate, and diffusion channel, for which the quasinormal modes are purely damped. This leads to a rather specific behavior of temperature
fluctuations. The transition from thermal to quantum
fluctuations is found in the region where hydrodynamics is barely applicable and is governed by the peculiar behavior of electromagnetic and gravitational scalar quasinormal modes.

As expected, temperature fluctuations behave classically for a large
black hole at high temperature and become more and more quantum as the black hole
temperature is lowered. For the large black holes the fluctuations are mostly given by 
the contribution from the sound mode, whereas for small black holes in quantum region 
the contributions from sound and diffusive modes are of the same order. 
Temperature fluctuations of a neutral AdS black
hole never become really quantum, because the black hole becomes
thermodynamically unstable and undergoes the Hawking-Page phase
transition when the fluctuations are about to enter the quantum regime. 
Temperature fluctuations of charged AdS black holes can, however,
enter the quantum regime. At low temperatures the behavior of the fluctuations is 
mostly governed by the lowest poles from the series of QNMs of spherical black holes. 
It would be interesting to study this region in more detail in future work.
At very low temperatures, for near-extremal black holes, fluctuations become 
so strong that one presumably cannot trust linear response theory any
more.  The strongly non-equilibrium behavior that then sets in can perhaps
be analyzed along the lines of
\cite{Sonner:2012if,Kundu:2013eba,Nakamura:2013yqa}. 

\subsection*{Acknowledgments}

We thank Joe Bhaseen, Elias Kiritsis, David Tong and Daniel Brattan
for useful comments and discussions. A.~K.~and S.~B.~G.~thank the
organizers of the HoloGrav network workshop ``Holographic Methods and
Applications'' for hospitality. 

The work of A.~B.~an Y.~K.~is supported by US DOE BES, ERC advanced grant DM
321-013 and by Swedish Research Council (VR) grant 2012-2983.  
The work of A.~K.~is partially supported by RFBR grant 12-02-00284-a and 
the Dynasty Foundation. The research of L.T.~is supported in part by Icelandic
Research Fund grant 130131-052 and by a grant from the University of Iceland Research Fund.
The work of K.Z.~is supported in part by the Marie
Curie network GATIS of the European Union's FP7 Programme under REA Grant
Agreement No 317089, in part by ERC advanced grant No 341222
and by Swedish Research Council (VR) grant 2013-4329.

\appendix

\section{Temperature fluctuations\label{app:A}}

\subsection{The overdamped mode}

\begin{figure}[!ht]
\begin{center}
\includegraphics[width=0.5\linewidth]{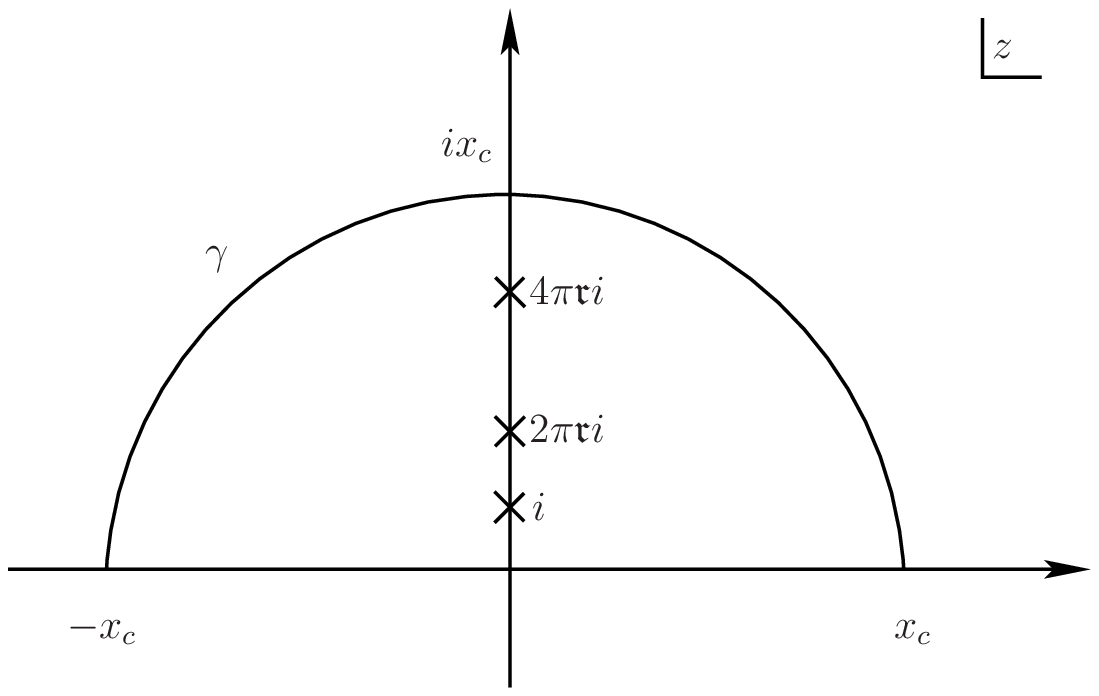}
\caption{Contour $\gamma$ in the complex $x$ plane.}
\label{fig:gamma1}
\end{center}
\end{figure}

Let us consider the integral
\beq
\langle\delta T^2\rangle = 
\frac{\hbar T}{2\pi C_v} \int_{-\infty}^{\infty} d\omega\;
\frac{\omega\tau}{(\omega\tau)^2 + 1}
\coth\left(\frac{\hbar\omega}{2T}\right)
=
\frac{\hbar T}{2\pi C_v\tau} \int_{-\infty}^{\infty} dx\;
\frac{x}{x^2 + 1}
\coth\left(\frac{x}{2\mathfrak{r}}\right),
\eeq
where $x=\omega\tau$ and we have defined
\beq
\mathfrak{r}\equiv \frac{T\tau}{\hbar}.
\eeq
We can now conveniently define
\beq
I\equiv \frac{2\pi C_v\tau}{\hbar T}\langle\delta T^2\rangle
= \int_{-\infty}^{\infty} dx\; f, \qquad
f \equiv \frac{x}{x^2 + 1} \coth\left(\frac{x}{2\mathfrak{r}}\right).
\eeq
As $I$ is UV-divergent, we introduce a cutoff $x_c\equiv\omega_c\tau$:
$I_{\rm reg} = \int_{-x_c}^{x_c} dx\; f$. 
Considering now the contour integral
\beq
I_\gamma = \oint_\gamma dz\;
\frac{z}{z^2 + 1} \coth\left(\frac{z}{2\mathfrak{r}}\right),
\eeq
where $\gamma$ is given in Fig.~\ref{fig:gamma1}.
The contribution from the arc will vanish for $x_c\to\infty$, i.e.
\beq
\lim_{x_c\to\infty} I_\gamma = 
\lim_{x_c\to\infty} {\rm p.v.}~I_{\rm reg},
\eeq
as long as the contour does not hit a pole on the imaginary axis. 

Using residue theory, we have
\beq
I_\gamma = 2\pi i \left[{\rm Res}(f,i)
+ \sum_{n=1}^{\Lambda} {\rm Res}(f,2n\pi\mathfrak{r}i)\right], \qquad
\Lambda \equiv {\rm floor}\left(\frac{x_c}{2\pi\mathfrak{r}}\right).
\eeq
The residues can be calculated easily
\begin{align}
{\rm Res}(f,i) &= -\frac{i}{2}\cot\left(\frac{1}{2\mathfrak{r}}\right),\\
{\rm Res}(f,2n\pi\mathfrak{r}i) &= 
\frac{4n\pi\mathfrak{r}^2i}{1 - 4n^2\pi^2\mathfrak{r}^2},
\end{align}
and thus the contour integral reads
\beq
I_\gamma = \pi\cot\left(\frac{1}{2\mathfrak{r}}\right)
-\sum_{n=1}^{\Lambda}
\frac{8n\pi^2\mathfrak{r}^2}{1 - 4n^2\pi^2\mathfrak{r}^2}.
\label{eq:contour_integral_result}
\eeq
So far we did not assume anything about $\mathfrak{r}$ other than it
being a real positive constant. 
In the limit $\mathfrak{r}\gg 2\pi$, we can expand the above
expression to get 
\beq
I_\gamma \simeq 2\pi\mathfrak{r} + 
2\sum_{n=1}^{\Lambda} \frac{1}{n},
\eeq
where the last term is the harmonic sum, which is known to be
divergent. Expressing the harmonic sum in terms of the cutoff, we have 
\beq
I_\gamma \simeq 2\pi\mathfrak{r} 
- \frac{\pi}{6\mathfrak{r}}
+ 2\log\left(\frac{x_c}{2\pi\mathfrak{r}}\right) 
+ 2\gamma_E 
+ \mathcal{O}\left(\frac{\mathfrak{r}}{x_c},\frac{1}{\mathfrak{r}^3}\right),
\eeq
where $\gamma_E$ is the Euler constant. 
Finally, we can write \cite{balatsky2003quantum} 
\beq
\langle\delta T^2\rangle = 
\frac{T^2}{C_v} \left[1
- \frac{\hbar^2}{12T^2\tau^2}
+ \frac{\hbar}{\pi T\tau}\left(
    \log\left(\frac{\hbar\omega_c}{2\pi T}\right)
    + \gamma_E\right) 
+ \mathcal{O}\left(\frac{1}{\omega_c\tau},\frac{\hbar^4}{(T\tau)^4}\right)
\right],
\eeq
where $\omega_c$ is the cut-off frequency.

In the opposite limit, $\mathfrak{r}\ll 2\pi$, the temperature
fluctuations are in the quantum regime and can be approximated as
follows. We neglect the residues coming from the response function and
approximate the sum in \eqref{eq:contour_integral_result} with an
integral and obtain 
\beq
I_\gamma \simeq \log x_c^2,
\eeq
which we can write as \cite{balatsky2003quantum} 
\beq
\langle\delta T^2\rangle \simeq
\frac{\hbar T}{\pi C_v\tau}\log\omega_c\tau.
\eeq

A more careful treatment of the arc is necessary because a simple
power counting naively tells us that it is logarithmically divergent. 
We want to show that
\beq
\lim_{x_c\to\infty} \int_{C_{x_c}^+} dz\;
\frac{z}{z^2 + 1} \coth\left(\frac{z}{2\mathfrak{r}}\right) = 0,
\eeq
where $C_{x_c}^+$ denotes the northern semicircle of 
radius $x_c$. The arc can be parametrized by $z=x_c e^{i\theta}$,
where $\theta\in[0,\pi]$. 
Rewriting the coth, we have
\beq
\coth\left(\frac{z}{2\mathfrak{r}}\right) = 
\left(1 + \frac{2e^{-z/\mathfrak{r}}}{1 - e^{-z/\mathfrak{r}}}\right) = 
\left(-1 - \frac{2e^{z/\mathfrak{r}}}{1 - e^{z/\mathfrak{r}}}\right),
\eeq
which we will use for the first and second quadrants of the $z$ plane,
respectively. We therefore have
\begin{align}
\int_{C_{x_c}^+} dz\;
\frac{z}{z^2 + 1} \coth\left(\frac{z}{2\mathfrak{r}}\right) &= 
\int_{C_{x_c}^{+1}} dz\;
\frac{z}{z^2 + 1} \left(1 + \frac{2e^{-z/\mathfrak{r}}}{1 - e^{-z/\mathfrak{r}}}\right)
+ \int_{C_{x_c}^{+2}} dz\;
\frac{z}{z^2 + 1} \left(-1 - \frac{2e^{z/\mathfrak{r}}}{1 - e^{z/\mathfrak{r}}}\right) \non
&= I_{+1}^{\rm pol} + I_{+1}^{\exp} + I_{+2}^{\rm pol} + I_{+2}^{\exp}.
\end{align}
The integrals of the two polynomials in $z$ can be carried out
\begin{align}
I_{+1}^{\rm pol} + I_{+2}^{\rm pol} &= 
i\int_0^{\frac{\pi}{2}} d\theta\; 
\frac{x_c^2 e^{i2\theta}}{x_c^2 e^{i2\theta} + 1}
-i\int_{\frac{\pi}{2}}^{\pi} d\theta\; 
\frac{x_c^2 e^{i2\theta}}{x_c^2 e^{i2\theta} + 1} 
= 2\Re\left[i\int_0^{\frac{\pi}{2}} d\theta\;
\frac{x_c^2 e^{i2\theta}}{x_c^2 e^{i2\theta} + 1}\right]
= -2\mathop{\rm arccoth}(x_c^2),
\end{align}
which goes to zero in the limit of $x_c\to\infty$. 
Now let us consider the first integral with exponentials
\beq
I_{+1}^{\exp} = 
\int_{C_{x_c}^{+1}} dz\;
\frac{z e^{-z/\mathfrak{r}}}{(z^2 + 1)(1 - e^{-z/\mathfrak{r}})}.
\eeq
Since we are interested in the large $x_c$ behavior of this integral,
let us instead consider
\beq
\tilde{I}_{+1}^{\exp} = 
\int_{C_{x_c}^{+1}} \frac{dz}{z}\;
\frac{e^{-z/\mathfrak{r}}}{1 - e^{-z/\mathfrak{r}}}
=
i \int_0^{\frac{\pi}{2}} d\theta \;
\frac{\exp\left(-\frac{x_c}{\mathfrak{r}}e^{i\theta}\right)}
{1 - \exp\left(-\frac{x_c}{\mathfrak{r}}e^{i\theta}\right)}.
\eeq
We will now need the lemma that
\beq
\left|\int_a^b dz\; f(z)\right| \leq \int_a^b dz\; M(z),
\eeq
where $f(z)$ is a complex function and $M(z)$ is a real-valued
function on the real interval $[a,b]$ such that $|f(z)|\leq M(z)$
everywhere on said interval (to prove this lemma, compose the integral 
into Riemann sums and use the triangle inequality). 
Hence we can write
\beq
|\tilde{I}_{+1}^{\exp}| \leq
\int_0^{\frac{\pi}{2}} d\theta \;
\frac{e^{-\frac{x_c}{\mathfrak{r}}\cos\theta}}
{\sqrt{1+e^{-\frac{2x_c}{\mathfrak{r}}\cos\theta} -
    2e^{-\frac{x_c}{\mathfrak{r}}\cos\theta}\cos\left(\frac{x_c}{\mathfrak{r}}\sin\theta\right)}}.
\label{eq:int_estimate}
\eeq
To simplify the problem we will now assume that
$x_c=(2m+1)\pi\mathfrak{r}$ with $m$ being an integer. 
This value of the cut off is chosen such that the contour goes right
in the middle between the Matsubara poles on the imaginary axis. 
We can now write
\beq
|\tilde{I}_{+1}^{\exp}| \leq
A\int_0^{\frac{\pi}{2}} d\theta \;
\exp\left(-\frac{x_c}{\mathfrak{r}}\cos\theta\right)
=
\frac{A\pi}{2}\left[I_0\big((2m+1)\pi\big) - L_0\big((2m+1)\pi\big)\right],
\eeq
where $A\gtrsim1$ is a constant of order one taking on the minimum
value of the denominator of eq.~\eqref{eq:int_estimate}. 
Expanding the above expression for large $m$, we have
\beq
|\tilde{I}_{+1}^{\exp}| \leq 
\frac{A}{(2m+1)\pi} + \mathcal{O}(m^{-3}),
\eeq
which clearly goes to zero for $m\to\infty$. 
The same proof can be applied to $I_{+2}^{\exp}$ and hence is bounded 
by the same numerical value.

\subsection{The underdamped mode}

\begin{figure}[!ht]
\begin{center}
\includegraphics[width=0.5\linewidth]{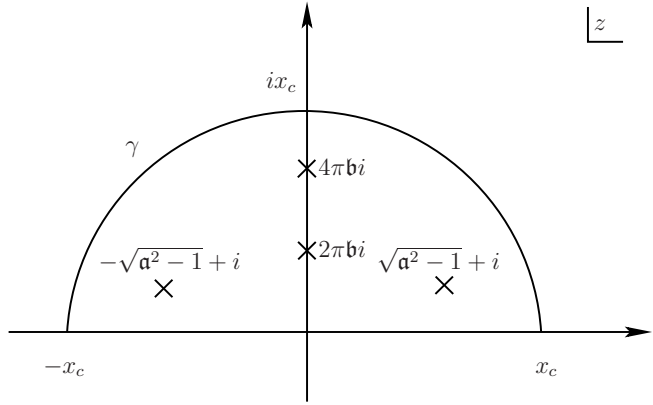}
\caption{Contour $\gamma$ in the complex plane.}
\label{fig:gamma2}
\end{center}
\end{figure}

Let us now consider the integral
\begin{equation}
\langle\delta T^2\rangle = 
\frac{\hbar T}{2\pi C_v} \int_{-\infty}^{\infty} d\omega\;
\frac{2\omega\Gamma\Omega^2}{(\omega^2-\Omega^2)^2 + 4\omega^2\Gamma^2}
\coth\left(\frac{\hbar\omega}{2T}\right)
=
\frac{\hbar T\Omega^2}{\pi C_v\Gamma}\int_{-\infty}^{\infty} dx\;
\frac{x}{(x^2-\mathfrak{a}^2)^2 + 4x^2}
\coth\left(\frac{x}{2\mathfrak{b}}\right),
\end{equation}
where we have defined
\beq
x \equiv \frac{\omega}{\Gamma}, \qquad
\mathfrak{a}\equiv \frac{\Omega}{\Gamma}, \qquad
\mathfrak{b}\equiv \frac{T}{\hbar\Gamma}.
\eeq
We will assume that $\mathfrak{a}>1$.
We can now conveniently define
\beq
I\equiv \frac{\pi C_v\Gamma}{\hbar T\Omega^2}\langle\delta T^2\rangle 
= \int_{-\infty}^{\infty} dx\; f, \qquad
f\equiv \frac{x}{(x^2-\mathfrak{a}^2)^2 + 4x^2}
\coth\left(\frac{x}{2\mathfrak{b}}\right). 
\eeq
Notice that for this response function, the integral is no longer
divergent. 
Consider now the contour integral
\beq
I_\gamma = \oint_\gamma dz\;
\frac{z}{(z^2-\mathfrak{a}^2)^2 + 4z^2} 
\coth\left(\frac{z}{2\mathfrak{b}}\right),
\eeq
where $\gamma$ is given in Fig.~\ref{fig:gamma2}.
The contribution from the arc will vanish for $x_c\to\infty$ and
$x_c\neq 2\pi\mathfrak{b}n$, meaning that the cutoff does not make the 
contour cut through a pole, and thus
\beq
\lim_{x_c\to\infty} I_\gamma = 
\lim_{x_c\to\infty} {\rm p.v.}~I.
\eeq
It is easy to estimate that the contribution from the arc will be of
the order of $1/x_c^2$.

We have two types of contributions, the first is due to the two
classical poles at $\pm\sqrt{\mathfrak{a}^2-1}+i$ and the second comes
from the $n$-th Matsubara mode at $2n\pi\mathfrak{b}i$,
\beq
I_\gamma = 2\pi i 
\left[\sum_{\pm}{\rm Res}(f,\pm\sqrt{\mathfrak{a}^2-1}+i) 
+ \sum_{n=1}^{\Lambda} {\rm Res}(f,2n\pi\mathfrak{b}i)\right], \qquad
\Lambda \equiv {\rm floor}\left(\frac{x_c}{2\pi\mathfrak{b}}\right).
\eeq
The residues can be calculated easily
\begin{align}
\sum_{\pm}{\rm Res}(f,\pm\sqrt{\mathfrak{a}^2-1}+i) &= 
-\frac{i}{8\sqrt{\mathfrak{a}^2-1}}\sum_{\pm}
\coth\left(\frac{\pm i+\sqrt{\mathfrak{a}^2-1}}{2\mathfrak{b}}\right),
\\
{\rm Res}(f,2n\pi\mathfrak{b}i) &= 
\frac{4n\pi\mathfrak{b}^2i}
{\left(4n^2\pi^2\mathfrak{b}^2+\mathfrak{a}^2\right)^2 
  - 16n^2\pi^2\mathfrak{b}^2},
\end{align}
and thus the contour integral reads
\beq
I_\gamma = 
\frac{\pi}{2\sqrt{\mathfrak{a}^2-1}}
\frac{\sinh\frac{\sqrt{\mathfrak{a}^2-1}}{\mathfrak{b}}}
{\cosh\frac{\sqrt{\mathfrak{a}^2-1}}{\mathfrak{b}}-\cos\frac{1}{\mathfrak{b}}}
-\sum_{n=1}^{\Lambda}
\frac{8n\pi^2\mathfrak{b}^2}
{\left(4n^2\pi^2\mathfrak{b}^2+\mathfrak{a}^2\right)^2 
  - 16n^2\pi^2\mathfrak{b}^2}.
\label{eq:cont_int}
\eeq

We will now consider the classical regime, which can be seen from
Fig.~\ref{fig:gamma2} to be where
\beq
2\pi\mathfrak{b} \gg \mathfrak{a}, \qquad {\rm or\ equivalently}\qquad
T \gg \frac{\hbar\Omega}{2\pi},
\eeq
which corresponds to the situation in which the classical poles are
reached at much lower frequencies (in the complex plane) than the
Matsubara modes. The classicality parameter (\ref{eq:q}) is given is
by $\mathfrak{q}=2\pi\mathfrak{b}/\mathfrak{a}$, so we can see
the how the classicality is manifested in the approximation. 
It will be instructive to expand the above contour integral for
$2\pi\mathfrak{b}\gg\mathfrak{a}>1$,
\beq
I_\gamma \simeq 
\frac{\pi\mathfrak{b}}{\mathfrak{a}^2}
+ \frac{\pi}{12\mathfrak{b}}
- \frac{\zeta(3)}{2\pi^2\mathfrak{b}^2}
+ \mathcal{O}\left(\frac{1}{\mathfrak{b}^3}\right),
\eeq
which we can write as
\beq
\langle\delta T^2\rangle \simeq
\frac{T^2}{C_v}\left[1
+ \frac{\hbar^2\Omega^2}{12T^2}
- \frac{\hbar^3\zeta(3)\Omega^2\Gamma}{2\pi^3T^3}
+ \mathcal{O}\left(\frac{\hbar^4\Omega^2\Gamma^2}{T^4}\right)
\right].
\eeq
If $\mathfrak{b}\ll\sqrt{\mathfrak{a}^2-1}$, then the contour integral
\eqref{eq:cont_int} can be approximated to be
\beq
I_\gamma \simeq \frac{\pi}{2\sqrt{\mathfrak{a}^2-1}}
\left[\frac{1}{2} +
\frac{1}{\pi}\arctan\left(\frac{\mathfrak{a}^2-2}{2\sqrt{\mathfrak{a}^2-1}}\right)
+ \mathcal{O}\left(\frac{\mathfrak{b}^2}{\mathfrak{a}^4}\right)
\right].
\eeq
If furthermore $\mathfrak{a}\gg 1$, the above expression simplifies as
follows
\beq
I_\gamma \simeq
\frac{\pi}{2\mathfrak{a}}
- \frac{1}{\mathfrak{a}^2}
+ \mathcal{O}\left(\frac{1}{\mathfrak{a}^3},\frac{\mathfrak{b}^2}{\mathfrak{a}^4}\right),
\eeq
which we can write as
\beq
\langle\delta T^2\rangle \simeq
\frac{\hbar T\Omega}{2C_v}
- \frac{\hbar T\Gamma}{\pi C_v}
+
T\Omega\mathcal{O}\left(\frac{\Gamma^2}{\Omega^2},\frac{T^2\Gamma}{\Omega^3}\right). 
\eeq
When $\mathfrak{a}<1$, the modes become those of the overdamped
regime. 

\section{Thermodynamics of the fluid with conserved charge \label{Thermo}}
In this Appendix we derive the thermodynamic identities used
throughout the text. Most of them may be looked up in
\cite{Kovtun:2012rj} or derived straightforwardly with the help of any
textbook on thermodynamics but we find it useful to write them here
for reference. 

In \cite{Kovtun:2012rj} one can find the following treatment. Consider
the change of the energy divided by the volume surrounding one
particle $d E /V$. In this case on has $V \sim n^{-1}$, $dV/V=-dn/n$
and one uses the thermodynamic equations valid in the grand canonical
ensemble 
\begin{equation}
\frac{dE}{V} = \frac{d (\epsilon V)}{V} = d \epsilon - \epsilon \frac{dn}{n}.
\end{equation}
On the other hand $dE = T dS - P dV$, hence
$\frac{T}{V}dS = d \epsilon - w \frac{dn}{n}$,
where $w$ is the enthalpy density. 
Then the variation of the pressure is
\begin{align}
dp = \frac{\p p}{\p \epsilon} \bigg|_{n} d \epsilon +  \frac{\p p}{\p n} \bigg|_{\epsilon} d n =  \left[\beta_1 + \frac{n}{w} \beta_2 \right] d \epsilon - \beta_2 \frac{n}{w} \frac{T dS}{V},
\end{align}
using the notation for the thermodynamic derivatives introduced in
eq.~\eqref{thermo_der}. 
Hence the speed of sound is
\begin{equation}
\label{speed-of-sound}
v_s^2 =  \left[\beta_1 + \frac{n}{w} \beta_2 \right] = \frac{\p p}{\p \epsilon} \bigg|_{S,N}\,.
\end{equation}

Let us introduce some more useful thermodynamic identities, using a
convenient formalism involving Jacobians 
\begin{align}
\frac{\p (A,B)}{\p(C,D)} \equiv \frac{\p A}{\p C} \bigg|_{D} \frac{\p B}{\p D} \bigg|_{C} - \frac{\p A}{\p D} \bigg|_{C} \frac{\p B}{\p C} \bigg|_{D}.
\end{align}
This allows us to rewrite
\begin{align}
T \left.\frac{\p \epsilon}{\p T} \right|_{\mu/T} =
T \frac{\frac{\p(\epsilon, \mu/T)}{\p(\mu, T)}}{\frac{\p(T, \mu/T)}{\p(\mu, T)}} 
= \mu \left.\frac{\p \epsilon}{\p \mu} \right|_{T} + T \left. \frac{\p
  \epsilon}{\p T} \right|_{\mu}.
\label{eq:TdepsilondT}
\end{align}
Similarly,
\begin{align}
 T \left.\frac{\p n}{\p T} \right|_{\mu/T} = \mu \left.\frac{\p n}{\p \mu} \right|_{T} + T \left. \frac{\p n}{\p T} \right|_{\mu}.
\end{align}
Using the identity $d \epsilon = T ds + \mu dn$, this can be rewritten
as 
\begin{align}
 T \left.\frac{\p n}{\p T} \right|_{\mu/T} = \left.\frac{\p
   \epsilon}{\p \mu} \right|_{T} - T \left[ \left.\frac{\p s}{\p \mu}
   \right|_{T} - \left. \frac{\p n}{\p T} \right|_{\mu} \right] =
 \left.\frac{\p \epsilon}{\p \mu} \right|_{T}.
\label{eq:TdndT}
\end{align}
The term in the square brackets is zero by a Maxwell relation,
because it may be expressed via the second derivatives of the
thermodynamic potential 
$d w = - s dT - n d\mu = -dp$ (because $w=-p$) in the grand canonical
ensemble: 
\begin{equation}
\label{pderiv}
s = \frac{\p p}{\p T} \bigg|_{\mu}, \qquad n =  \frac{\p p}{\p \mu} \bigg|_{T}.
\end{equation}
Using eqs.~\eqref{eq:TdepsilondT} and \eqref{eq:TdndT} we can
reexpress the coefficients, $\alpha$, from (\ref{thermo_der}) as
\begin{align}
\alpha_1 &= 
T \frac{\p(\mu/T)}{\p \epsilon} \bigg|_{n} =
T \frac{\frac{\p(\mu/T, n)}{\p(\mu/T,  T)}}{\frac{\p(\epsilon, n)}{\p(\mu/T,  T)}}= 
\frac{1}{T D} \frac{\p \epsilon}{\p \mu} \bigg|_{T}, \\
\alpha_2 &=
T \frac{\p(\mu/T)}{\p n} \bigg|_{\epsilon} =
T \frac{\frac{\p(\mu/T, \epsilon)}{\p(\mu/T,  T)}}{\frac{\p(n, \epsilon)}{\p(\mu/T,  T)}}= 
- \frac{1}{D} \left[ \frac{\p \epsilon}{\p T} \bigg|_{\mu} + \frac{\mu}{T} \frac{\p \epsilon}{\p \mu} \bigg|_{T}  \right],
\end{align}
where we again have used the Maxwell relation and $D$ in the
denominator is given by
\begin{align}
D = \frac{\p \epsilon}{\p \mu} \bigg|_{T} \frac{\p n}{\p T} \bigg|_{\mu/T} -  \frac{\p n}{\p \mu} \bigg|_{T} \frac{\p \epsilon}{\p T} \bigg|_{\mu/T} &= 
\frac {1}{T} \frac{\p \epsilon}{\p \mu} \bigg|_{T} \frac{\p \epsilon}{\p \mu} \bigg|_{T} -  \frac{\p n}{\p \mu} \bigg|_{T} \left( \frac{\mu}{T} \frac{\p \epsilon}{\p \mu} \bigg|_{T} + \frac{\p \epsilon}{\p T} \bigg|_{\mu} \right) =
\non
& = \frac {1}{T} \frac{\p \epsilon}{\p \mu} \bigg|_{T}  T \frac{\p s}{\p \mu} \bigg|_{T}  - \frac{\p n}{\p \mu} \bigg|_{T} \frac{\p \epsilon}{\p T} \bigg|_{\mu} = 
\frac{\p(\epsilon, n)}{\p(\mu , T)}. \label{det}
\end{align}
To proceed with the coefficients, $\beta$, defined in
eq.~\eqref{thermo_der2} we use (\ref{pderiv}) 
\begin{align}
\beta_1 &= \frac{\p p}{\p \epsilon} \bigg|_{n} = \frac{1}{D} \frac{\p(p, n)}{\p (\mu, T)} = \frac{1}{D} \left( n \frac{\p s}{\p \mu} \bigg|_{T} - s \frac{\p n}{\p \mu} \bigg|_{T} \right), \\
\beta_2 &= \frac{\p p}{\p n} \bigg|_{\epsilon} = - \frac{1}{D} \frac{\p(p, \epsilon)}{\p (\mu, T)} = - \frac{1}{D} \left( n \frac{\p \epsilon}{\p T} \bigg|_{\mu} - s \frac{\p \epsilon}{\p \mu} \bigg|_{T} \right).
\end{align}
Now one can show that
\begin{align}
\label{off-diag}
w \alpha_1 + n \alpha_2 &= -\frac{1}{D} \left[ - \frac{\epsilon + p}{T} \frac{\p \epsilon}{\p \mu} \bigg|_{T} + \frac{n \mu}{T} \frac{\p \epsilon}{\p \mu} \bigg|_{T} + n \frac{\p \epsilon}{\p T} \bigg|_{\mu} \right] = \beta_2,\\
n \alpha_1 - \beta_1 &= \frac{1}{D} \left( \frac{n \mu}{T} \frac{\p n}{\p \mu} \bigg|_{T} + s \frac{\p n}{\p \mu} \bigg|_{T} \right) = \frac{w}{D T} \frac{\p n}{\p \mu} \bigg|_{T} \equiv - w \alpha_3,
\end{align}
and in turn it follows that
$\alpha_2\beta_1-\alpha_1\beta_2=w(\alpha_2\alpha_3-\alpha_1^2)$,
$n\alpha_2-\beta_2=w\alpha_1$ from which we can write
eq.~\eqref{dT_classical}. We have introduced the quantity
$\alpha_3$ with which one can show that the following holds 
\begin{align}
\alpha_1^2 - \alpha_2 \alpha_3 
&= \frac{1}{D^2 T} \left[ \frac{\p \epsilon}{\p \mu} \bigg|_{T} \left(\frac{1}{T}  \frac{\p \epsilon}{\p \mu} \bigg|_{T} - \frac{\mu}{T} \frac{\p n}{\p \mu} \bigg|_{T} \right) -  \frac{\p \epsilon}{\p T} \bigg|_{\mu} \frac{\p n}{\p \mu} \bigg|_{T} \right] =  \frac{1}{D^2 T} \frac{\p(\epsilon, n)}{\p(\mu , T)} = \frac{1}{D
  T}, \label{alphas}
\end{align}
where we used the same tricks as in (\ref{det}).
Using Jacobians, it can be seen that
$$
\alpha_3=\frac{1}{DT}\left.\frac{\p n}{\p\mu}\right|_T
=\frac{1}{T}\left.\frac{\p T}{\p \epsilon}\right|_n,
$$
as indicated in the main text. 

When calculating the correlator of temperature fluctuations we encounter the following expressions
\begin{align}
\label{temp_derivatives}
\frac{\p T}{\p \epsilon} \bigg|_{n} &=  \frac{\p(T, n)}{\p(\epsilon , n)} = - \frac{1}{D} \frac{\p(T, n)}{\p(T , \mu)} = -\frac{1}{D} \frac{\p n}{\p \mu} \bigg|_{T}, \\
\frac{\p T}{\p n} \bigg|_{\epsilon} &=  \frac{\p(T, \epsilon)}{\p(n,\epsilon)} = \frac{1}{D} \frac{\p(T, \epsilon)}{\p(T , \mu)} = \frac{1}{D} \frac{\p \epsilon}{\p \mu} \bigg|_{T}.
\end{align}
Hence, we can rewrite the expression one finds in the numerator of the
fluctuation amplitude in classical limit (\ref{dT_classical}) as 
\begin{align}
&\left( \frac{\p T}{\p \epsilon} \bigg|_{n} \right)^2 \alpha_2 - 2  \frac{\p T}{\p \epsilon} \bigg|_{n} \frac{\p T}{\p n} \bigg|_{\epsilon} \alpha_1 + \left( \frac{\p T}{\p n} \bigg|_{\epsilon}  \right)^2 \alpha_3 = 
\non
&= - \frac{1}{D^3} \frac{\p n}{\p \mu} \bigg|_{T} \left( \frac{\p
    n}{\p \mu} \bigg|_{T} \frac{\p \epsilon}{\p T} \bigg|_{\mu} +
  \frac{\p n}{\p \mu} \bigg|_{T}  \frac{\mu}{T} \frac{\p \epsilon}{\p
    \mu} \bigg|_{T} - \frac{1}{T} \frac{\p \epsilon}{\p \mu}
  \bigg|_{T}   \right) = -\frac{1}{D^2} \frac{\p n}{\p \mu} \bigg|_{T}
  = \frac{1}{D} \frac{\p T}{\p \epsilon} \bigg|_{n} = \frac{1}{D
    c_v}. \label{temperature_numerator}
\end{align}
In the last equation the volumetric heat capacity is defined at
constant self-volume of one particle, which means $V \sim 1/n = {\rm const}$.
Finally, \eqref{dT_classical} is obtained by combining \eqref{alphas} and \eqref{temperature_numerator}.

\section{Calculation of the scalar quasinormal modes of an 
AdS-Reissner-Nordstr\"{o}m black hole \label{app:B}}

\subsection{Boundary conditions}

The equations of motion, which describe the coupled fluctuations of
the metric and electromagnetic field on the background of a
Reissner-Nordstr\"om black hole in 4-dimensional de Sitter space were obtained
in \cite{Mellor:1989ac} following the procedure outlined in
\cite{chandrasekhar1998mathematical}. The equations of motion for the
AdS case can be obtained simply by considering a negative cosmological
constant $\Lambda = - \frac{3}{R^2}$. The notation used in this
section differs from the one used in the rest of the paper, but we adopt
it in order to facilitate reference to
\cite{Mellor:1989ac,chandrasekhar1998mathematical} keeping in mind
that the final result (quasinormal frequencies) will be easy to
interpret. 

The background metric is
\begin{equation}
\label{metric}
ds^2 = e^{2 \nu} dt^2 - e^{2 \psi} d\phi^2 - e^{2 \mu_2} dr^2 - e^{2 \mu_3} d\theta^2, 
\end{equation}
where
\begin{equation}
e^{2 \nu} = \frac{\Delta}{r^2}, \quad  e^{2 \mu_2} =
\frac{r^2}{\Delta}, \quad e^{2 \mu_3} = r^2, \quad e^{2 \psi} = r^2
\sin^2 \theta , 
\end{equation}
and 
\begin{equation}
\Delta = r^2 - 2 Mr + Q^2 + \frac{r^4}{R^2}.
\end{equation}
The background electromagnetic field is described by a single
component of the field-strength tensor 
\begin{equation}
F_{t r} = - \frac{Q}{r^2}.
\end{equation}
The polar (even with respect to the change of sign of $\phi$) mode
involves the fluctuations of the metric $\delta \nu, \delta \mu_2,
\delta \mu_3, \delta \psi$ and of the field-strength tensor $\delta
F_{tr}, F_{t \theta}, F_{r \theta}$. The fluctuations with angular
momentum $l$ and the frequency $\omega$ may be described in the
following way  
\begin{align}
\label{dN}
\delta \nu &= N(r) P_l (\theta), & \delta F_{tr} &= \frac{r^2 e^{2 \nu}}{2 Q} B_{tr}(r) P_{l}, \\
\label{dL}
\delta \mu_2 &= L(r) P_l (\theta), & F_{t \theta} &= - \frac{r e^\nu}{2 Q} B_{t \theta} P_{l,\theta}, \\
\label{dBr}
\delta \mu_3 &= T(r) P_l + V(r) P_{l,\theta, \theta}, & F_{r \theta} &= - \frac{i \omega r e^{-\nu}}{2 Q} B_{r \theta}(r) P_{l, \theta}, \\
\notag
\delta \psi &= T(r) P_l + V(r) P_{l,\theta} \cot \theta . 
\end{align}
Introducing the parameters
\begin{align}
\mu^2 &= 2n = (l-1)(l+2), \\
\label{p1p2}
p_1 &= 3M + (9 M^2 + 4 Q^2 \mu^2)^{1/2}, \\
\notag
p_2 &= 3M - (9 M^2 + 4 Q^2 \mu^2)^{1/2},
\end{align}
one can rewrite the Einstein equations for fluctuations
\begin{equation}
\delta R_{a b} = -2 \left[ \eta^{nm}(\delta F_{a n} F_{b m} + F_{an} \delta F_{bm}) - \eta_{a b} Q \delta F_{tr}/r^2 \right] 
\end{equation}
as a system of differential equations of first order \cite{Mellor:1989ac}
\begin{align}
\label{Nr}
&N_{,r} = aN + bL + c(n V - B_{r \theta}), \\
\label{Lr}
&L_{,r} = \left(a - \frac{1}{r} + \nu_{,r} \right) N + \left(b - \frac{1}{r} - \nu_{,r} \right)L + c (n V - B_{r \theta}) - \frac{2}{r} B_{r \theta}, \\
\label{Vr}
&n V_{,r} = - \left(a - \frac{1}{r} + \nu_{,r} \right) N - \left(b + \frac{1}{r} - 2 \nu_{,r} \right)L - \left(c + \frac{1}{r} - \nu_{,r} \right) (n V - B_{r \theta}) + B_{t \theta}, \\
\label{Bt}
&B_{t \theta} = {B_{r \theta}}_{,r} + \frac{2}{r} B_{r \theta}, \\
&r^4 e^{2 \nu} B_{tr} = 2 Q^2 \left[2 T - l(l+1)V \right] - l (l+1)r^2 B_{r \theta}, \\
\label{Max}
&(r^2 e^{2 \nu} B_{t \theta})_{,r}  + r^2 e^{2\nu} B_{tr} + \omega^2 r^2 e^{-2 \nu} B_{r \theta} = 2 Q^2 \frac{N+L}{r},
\end{align}
where
\begin{align}
\label{a}
a &= \frac{n+1}{r} e^{-2 \nu}, \\
\label{b}
b &= - \frac{1}{r} + \nu_{,r} + r \nu^2_{,r} + \omega^2 e^{-4 \nu} r - 2 \frac{e^{-2 \nu}}{r^3} Q^2 - \frac{n e^{- 2 \nu}}{r}, \\
\label{c}
c &= - \frac{1}{r} + r \nu^2_{,r} + \omega^2 e^{-4 \nu} r - \frac{2 e^{-2 \nu}}{r^3} Q^2 + \frac{e^{-2 \nu}}{r}.
\end{align}
These equations can be decoupled upon introducing the functions
\begin{align}
\label{Z1}
Z_1^+ &= p_1 H_1^+ + (-p_1 p_2)^{1/2} H_2^+,\\
\label{Z2}
Z_2^+ &= -(-p_1 p_2)^{1/2} H_1^+ + p_1 H_2^+,
\end{align}
where
\begin{align}
H_1^+ &= - \frac{1}{Q \mu} \left[r^2 B_{r \theta} + 2 Q^2 \frac{r}{\bar{\omega}}(L + nV - B_{r \theta}) \right], \\
H_2^+ &= rV - \frac{r^2}{\bar{\omega}} (L + nV - B_{r \theta}),
\end{align}
and
\begin{equation}
\bar{\omega} = nr + 3M - \frac{2 Q^2}{r}. 
\end{equation}
The system can be reduced to the couple of equations
\begin{equation}
\label{shr}
\frac{\Delta}{r^2} \frac{d}{dr}\left(\frac{\Delta}{r^2} \frac{d}{dr} Z_i^+ \right) + \omega^2 Z_i^+ = V_i^+ Z_i^+ \qquad (i=1,2) 
\end{equation}
with Schr\"{o}dinger-type potentials
\begin{align}
\label{potentials}
V_1^+ &= \frac{\Delta}{r^5} \left[U + \frac{1}{2}(p_1 - p_2) W \right], \\
V_2^+ &= \frac{\Delta}{r^5} \left[U - \frac{1}{2}(p_1 - p_2) W \right],
\end{align}
where
\begin{align}
U &= \left(2 n r + 3 M \right) W + \left( \bar{\omega} - nr - M + 2 \frac{r^3}{R^2} \right) - \frac{2 n r^2}{\bar{\omega}} e^{2 \nu}, \\
W &= \frac{\Delta}{r \bar{\omega}^2}(2 n r + 3M) + \frac{1}{\bar{\omega}} \left(nr + M -2 \frac{r^3}{R^2} \right).
\end{align}

At this point it is useful to note, that for vanishing charge ($Q \rar
0$) the potential $V_2^+$ reduces to the potential for purely
gravitational polar fluctuations of the Schwarzschild black hole in AdS
\cite{Berti:2003ud}. It is not surprising, because in this limit $p_2$
vanishes and hence $Z_2^+$ reduces to purely gravitational
fluctuations $H_2^+$ plus a term proportional to $B_{r \theta}$, which
vanishes itself according to the definition $B_{r \theta} \sim Q F_{r
  \theta}$ (\ref{dBr}). Similarly, in this limit the mode described by
$Z_1^+$ corresponds to purely electromagnetic fluctuations on the
background of the neutral black hole. Keeping these connections in mind at
nonzero $Q$, we will still call the modes associated with $Z_1^+$ and
$Z_2^+$  ``electromagnetic'' and ``gravitational'', respectively. 

In what follows, we will consider the fluctuations described by $Z_i^+$
and recover the asymptotic behavior of the metric fields in this
mode. In order to do this, we need to complete the solution of
(\ref{Nr}-\ref{Max}) following the procedure outlined in
\cite{chandrasekhar1998mathematical}. First of all, we note that the
sum of (\ref{Lr}) and (\ref{Vr}) after the substitution of (\ref{Bt})
may be written as 
\begin{equation}
\label{LX}
L_{,r} + \left(\frac{2}{r} - \nu_{,r} \right) L = - \left[ X_{,r} + \left(\frac{1}{r} - \nu_{,r} \right) X \right], 
\end{equation}
where
\begin{equation}
\label{X}
X = n V - B_{r \theta}. 
\end{equation}
On the other hand, we notice that the linear combination of $H_1^+$ and $H_2^+$, which we denote by $Z^*$, is expressed in terms of $L$ and $X$ as
\begin{align}
\label{Zstar}
Z^* =  n H_2^+ + \frac{Q \mu}{r} H_1^+  &= \frac{1}{\bom} (3 M r - 4 Q^2)  X - \frac{1}{\bom} (n r^2 + 2 Q^2) L \\
\notag
& = r X - \frac{n r^2 + 2 Q^2}{\bom} (L + X) .
\end{align}
Substituting $X$ from this expression into (\ref{LX}) we get
\begin{equation}
\bom \frac{d}{dr} \left( \frac{r^3 e^{-\nu}}{3 M r - 4 Q^2} L \right) = - r \frac{d}{dr} \left(r e^{-\nu} \frac{\bom}{3 M r - 4 Q^2} Z^* \right).
\end{equation}
The expression for $L$ is thus obtained via the integral
\begin{align}
\label{LZ}
L &= -  \frac{3 M r - 4 Q^2}{r^3 e^{-\nu}} \int dr \frac{r}{\bom} \frac{d}{dr} \left(r e^{-\nu} \frac{\bom}{3 M r - 4 Q^2} Z^* \right) \\
\notag
 &= - \frac{1}{r} Z^* +  \frac{3 M r - 4 Q^2}{r^3 e^{-\nu}} \left[\int dr \frac{e^{-\nu}}{\bom} Z^* + C \right], 
\end{align}
where in the second line we have performed an integration by parts and
$C$ is an integration constant. 
However, the expression for $X$ can similarly be obtained by
substituting $L$ of (\ref{Zstar}) into (\ref{LX}). 
After the integration by parts it reads
\begin{equation}
\label{XZ}
X  = \frac{1}{r} Z^* + \frac{n r^2 + 2 Q^2}{r^3 e^{-\nu}} \left[ \int dr \frac{e^{-\nu}}{\bom} Z^* + C \right].  
\end{equation}
One can check, that the constants of integration in (\ref{LZ}) and
(\ref{XZ}) are consistent by plugging these expressions into
(\ref{LX}). We notice that the sum of $L$ and $X$ assumes a concise
form 
\begin{equation}
\label{LXZ}
L + X =  \frac{\bom}{r^2 e^{-\nu}} \left[ \int dr \frac{e^{-\nu}}{\bom} Z^* + C \right]. 
\end{equation}
To proceed with the evaluation of $N$, we take the derivative (\ref{Zstar}) and substitute the expression of $(L + X)_{,r}$ from (\ref{LX})
\begin{equation}
Z^*_{,r} = r X_{,r} + \left(\frac{3 M r - 4 Q^2}{r \bom} \right) X -
r^2 e^{-\nu} \frac{d}{dr} \left(\frac{1}{r^2 e^{-\nu}} \frac{n r^2 + 2
  Q^2}{\bom}  \right) (L+X) .
\end{equation}
Finally, we use equations (\ref{Vr}) and (\ref{Bt}) to eliminate
$X_{,r}$ and obtain the expression for $N$ 
\begin{align}
\label{NZ}
N &= \frac{1}{r a - 1 + r \nu_{,r}} \bigg \{\frac{2}{r} B_{r \theta} - \frac{d}{dr} Z^*  -  \\
\notag
& - \left[rb + 1 - 2 r \nu_{,r} + r^2 e^{-\nu} \frac{d}{dr} \left(\frac{1}{r^2 e^{-\nu}} \frac{n r^2 + 2 Q^2}{\bom}  \right) \right] \frac{\bom}{r^2 e^{-\nu}} \left( \int dr \frac{e^{-\nu}}{\bom} Z^* + C \right) + \\
\notag
& + \left(\frac{3 M r - 4 Q^2}{r \bom} - a r \right)  \left[\frac{1}{r} Z^* + \frac{n r^2 + 2 Q^2}{r^3 e^{-\nu}} \left( \int dr \frac{e^{-\nu}}{\bom} Z^* + C \right)  \right]  \bigg \}.
\end{align}

Let us now turn to the asymptotic behavior of the master functions $Z_1^+$ and $Z_2^+$. The corresponding Schr\"odinger equations at large $r$ take the form
\begin{align}
\left( \frac{r^2}{R^4} \p_{r} r^2 \p_r + \omega^2 - \frac{\mu^2 + 2}{R^2} - \frac{2 p_1^2}{\mu^4 R^4} \right) Z_2^+  &= 0, \\
\left( \frac{r^2}{R^4} \p_{r} r^2 \p_r + \omega^2 - \frac{\mu^2 + 2}{R^2} - \frac{2 p_2^2}{\mu^4 R^4} \right) Z_1^+  &= 0.
\end{align}
As expected in the limit $Q \rar 0$ the equation for $Z_2^+$ reduces
to that of the Schwarzschild black hole in
\cite{Michalogiorgakis:2006jc}. Hence, for large $r$, the asymptotics 
of the master functions can be expressed as sums of linearly
independent modes 
\begin{align*}
Z_2^+ \Big|_{r \rar \infty} &= \alpha + \frac{\beta}{r}, \\
Z_1^+ \Big|_{r \rar \infty} &= \gamma + \frac{\delta}{r}.
\end{align*}
From (\ref{Z1}), (\ref{Z2}) we get the expressions for $H_1^+$ and $H_2^+$
and from (\ref{Zstar}) the asymptotic expression for $Z^*$ 
\begin{align}
\label{asZ}
Z^* = \xi + \frac{\eta}{r} + \frac{\zeta}{r^2},
\end{align}
where
\begin{align}
\label{xis}
\xi &= \frac{n \big(p_1 \alpha + \sqrt{-p_1 p_2} \ \gamma\big)}{p_1 (p_1 - p_2)}, \\
\eta &= \frac{n \big(p_1 \beta + \sqrt{-p_1 p_2} \ \delta \big) + Q \mu \big( p_1 \gamma - \sqrt{-p_1 p_2} \ \alpha \big)}{p_1 (p_1 - p_2)}, \\
\zeta &= \frac{Q \mu \big( p_1 \delta - \sqrt{-p_1 p_2} \ \beta \big)}{p_1 (p_1 - p_2)}. 
\end{align}
After plugging this expansion into (\ref{LZ}), (\ref{XZ}) and (\ref{NZ}) we get at large $r$
\begin{align}
L \Big|_{r \rar \infty} &= - \frac{\xi - \frac{3 M}{R} C }{r} -
\frac{\eta + \frac{3 M}{n} \xi + 4 C \frac{Q^2}{R}}{r^2} +
O\left(\frac{1}{r^3}\right) , \\
V \Big|_{r \rar \infty} &= \frac{C}{R} + \frac{ \eta + \frac{3 M}{n}
  \xi  +  C \left( 4 \frac{Q^2}{R} + n \right)}{2 n r^2} +
O\left(\frac{1}{r^3}\right) , \\
N \Big|_{r \rar \infty} &= - \frac{ \eta + \frac{3 M}{n} \xi  -  C n
  R^3 \omega^2}{r^2} + O\left(\frac{1}{r^3}\right) .
\end{align}
We should note here that because $F_{r \theta}$ behaves as $r^{-2}$ on
the boundary (one can see this from the $r\rightarrow\infty$ expansion of the
Maxwell equations) the function $B_{r \theta}$ which enters the
definition of $X$ (\ref{X}) and the expression for $N$ (\ref{NZ})
falls off as $r^{-3}$ and does not enter the above expansions.
As discussed in \cite{Michalogiorgakis:2006jc}, the
perturbations of the metric near the AdS boundary have two linearly
independent modes, which behave as $r^2$ and $\frac{1}{r}$. The former
violates the asymptotic behavior of the AdS metric on the boundary
and should be forbidden. The latter can be nicely interpreted in
the AdS/CFT correspondence as a vacuum expectation value of the
stress-energy tensor of the dual field theory and thus we need to keep
it. According to the definitions (\ref{metric}), (\ref{dN}--\ref{dBr})
keeping the mode $\sim r^{-1}$ in the boundary metric 
fluctuations, means keeping only the mode $\sim r^{-3}$ in the
functions $N$ and $V$. Hence, we need to choose the constant of
integration $C$ equal to zero and demand the Robin boundary conditions
on the wave function $Z^*$ 
\begin{equation}
 \eta = - \frac{3 M}{n} \xi.
\end{equation}
Taking into account the definitions (\ref{xis}) we can derive the
conditions on the master wave functions $Z_1^+$ and $Z_2^+$. Because
the equations (\ref{shr}) are independent we can consider
the ``gravitational'' and ``electromagnetic'' modes separately. Thus we
find  
\begin{align}
\label{bcB}
\beta &= - \left(\frac{3 M}{n} + \frac{4 Q^2 }{p_1} \right) \alpha, && \mbox{when } Z_1^+ = 0; &&\\ 
\delta &= -\left(\frac{3 M}{n} - \frac{p_1}{2 n} \right) \gamma, && \mbox{when } Z_2^+ = 0. &&
\end{align}
We note that to obtain this result one needs to take the negative
branch of the square root in (\ref{p1p2}): $\sqrt{4 \mu^2 Q^2} = - 2
\mu Q$. This choice is motivated by the comparison with the hydrodynamic
treatment \eqref{eq:hydro}, discussed previously. Taking the positive
branch would give results which are inconsistent with
hydrodynamics, so we ignore this possibility as unphysical. 
One can check, that in the limit $Q \rar 0$ the first of these
conditions coincides with that obtained for the Schwarzschild black hole in
\cite{Michalogiorgakis:2006jc}. This is consistent with the fact
pointed out earlier, that the $Z_2^+$ mode becomes purely gravitational in
this limit. The second condition vanishes in this case, because at
$Q=0$ the ``electromagnetic'' mode does not couple to gravity and the
treatment based on the asymptotic behavior of the metric is no longer
valid. 

The boundary conditions at the horizon $r= r_+$ are easier to
obtain. By definition, the quasinormal mode should contain only the
wave ``infalling'' to the horizon. In ``tortoise'' coordinates $dr^* =
\frac{r^2}{\Delta} dr$ the Schr\"{o}dinger equation (\ref{shr}) takes
the simple form 
\begin{equation}
\left[ \p_{r_*}^2 - \p_{\tau}^2 - V_i^+ \right] Z_i^+ = 0, \quad (i=1,2) 
\end{equation}
Noticing that  $V_i^+$ vanishes at the horizon, we get the
infalling wave solution in the form 
\begin{equation}
\label{bcH}
Z_i^+ \Big|_{r_* \rar -\infty} \sim e^{-i \omega(\tau  + r_*)}.
\end{equation}

\subsection{Numerical solution}
In order to proceed with the numerical calculation of the quasinormal
modes, we make several redefinitions of variables. First of all we
substitute the infalling wave Ansatz 
\begin{align}
\label{psi}
Z =  e^{-i \omega(\tau  + r_*)} \psi(r) ,
\end{align}
and get the equation for $\psi$
\begin{equation}
\label{psiequ}
\psi''(r) + \left[\frac{r^2}{\Delta} \frac{d}{dr}  \frac{\Delta}{r^2}
  - 2 i \omega  \frac{r^2}{\Delta} \right]  \psi'(r) -
\frac{r^4}{\Delta^2} V_i^+ \psi(r) = 0 .
\end{equation}
To compactify the interval of integration we introduce the variable
$y= 1 - \frac{r_+}{r}$. After this substitution the boundary of AdS is
located at $y=1$ and the horizon is at $y=0$. The boundary conditions
for $\psi(y)$ can be easily derived from (\ref{bcB}) and
(\ref{bcH}). At the horizon the infalling wave boundary condition
is simply stated as
\begin{equation}
\label{psibcH}
\psi(0)=1.
\end{equation}
On the AdS boundary, the condition is found from the expansion of
(\ref{psi}) at $r \rar \infty$,
\begin{align}
\label{psibcB}
\psi(y) \Big|_{y\rar1} &= 1 + \frac{1}{r_+} \left( \frac{3 M}{n} +
\frac{4 Q^2 }{p_1} + i \omega \right) (y-1) + \dots & &\mbox{for the ``gravitational'' mode,} \\
\psi(y) \Big|_{y\rar1} &= 1 + \frac{1}{r_+} \left( \frac{3 M}{n} -
\frac{p_1}{2 n} + i \omega \right) (y-1) + \dots & &\mbox{for the ``electromagnetic'' mode.}
\end{align}
Similarly to \cite{Michalogiorgakis:2006jc} we use these boundary conditions 
to expand the solution in a series around the singular points of (\ref{psiequ}) at
$y=0$ and $y=1$ to sufficiently high order and then solve the equation numerically
by seeding the shooting procedure from both ends of the interval. We then look for
a frequency $\omega_0$ at which the Wronskian of the two solutions
coming from opposite ends is zero at an intermediate point.  This
tells us that at that given frequency, the shooting solutions can be smoothly connected,
resulting in a nontrivial solution to (\ref{psiequ}) on the full interval with boundary
conditions (\ref{psibcH},\ref{psibcB}). This solution is the
quasinormal mode and the frequency $\omega_0$ is the quasinormal
frequency of the black hole.


\end{document}